\documentclass[acmsmall,screen,nonacm]{acmart}

\usepackage[utf8]{inputenc}

\usepackage{enumitem}
\usepackage{algorithm,algpseudocode}
\usepackage{amsmath,amssymb,amsthm,mathtools,mathpartir}
\usepackage{stmaryrd}
\usepackage{booktabs,array,longtable}
\usepackage{float}
\usepackage{placeins}
\usepackage{wrapfig}
\usepackage{xcolor}
\usepackage{colortbl}
\usepackage{listings}
\usepackage{tikz}
\usetikzlibrary{arrows.meta,calc}
\usepackage{subcaption}
\usepackage[capitalize]{cleveref}

\crefname{listing}{Listing}{Listings}
\creflabelformat{listing}{#2(#1)#3}
\newenvironment{listingdisplay}[1]{\crefalias{equation}{listing}\equation\label{#1}}{\endequation}

\DeclareFontFamily{U}{stmry}{}
\DeclareFontShape{U}{stmry}{m}{n}{%
  <-6> stmary5
  <6-7> stmary6
  <7-8> stmary7
  <8-9> stmary8
  <9-10> stmary9
  <10-> stmary10
}{}

\newcommand{\sys}{\textsc{Isomer}}

\newcommand{\code}[1]{\texttt{\detokenize{#1}}}

\newcommand{\Aref}[1]{Appendix~\ref{#1}}

\newcommand{\Un}{\textstyle\bigcup\nolimits}
\newcommand{\Du}{\textstyle\bigsqcup\nolimits}
\newcommand{\Sum}[1]{\Un\{#1\}}
\newcommand{\Tagged}[2]{\Du^{#1}\{#2\}}
\newcommand{\ar}[2]{#1\Rightarrow #2}
\newcommand{\Uint}[1]{\mathsf{u}_{#1}}
\newcommand{\RefT}[1]{\mathsf{#1}}
\newcommand{\RefL}[1]{\mathsf{ref}\,\RefT{#1}}
\newcommand{\Stored}[1]{\langle #1\rangle}
\newcommand{\Pad}[1]{\mathsf{pad}_{#1}}
\newcommand{\Zero}[1]{\mathsf{zero}_{#1}}
\newcommand{\Ptr}[2]{\&_{#1}(#2)}
\newcommand{\Ctor}[1]{\mathbf{#1}}

\newcommand{\IR}{\mathbb{S}}
\newcommand{\cons}[2]{\mathsf{cons}_{#1}(#2)}
\newcommand{\smallunit}{\raisebox{.25ex}{$\scriptstyle()$}}
\newcommand{\PtrVal}[2]{\mathsf{ptr}(#1,#2)}
\newcommand{\W}{\mathcal W}
\newcommand{\Ctx}{\mathcal{C}}

\newcommand{\upsdWaux}[2]{\rotatebox[origin=c]{180}{{\normalfont\unboldmath$#1\W$}}}
\newcommand{\upsdW}{\mathord{\mathpalette\upsdWaux{}}}

\theoremstyle{acmplain}
\newtheorem{theorem}{Theorem}
\newtheorem{lemma}[theorem]{Lemma}
\theoremstyle{acmdefinition}
\newtheorem{definition}[theorem]{Definition}

\lstdefinelanguage{sys}{
  sensitive=true,
  morekeywords={data,fn,fun,let,in,match,if,then,else,as,io,inl,inr},
  morecomment=[l]{//},
  morestring=[b]"
}

\lstdefinestyle{sys}{
  language=sys,
  basicstyle=\small\ttfamily,
  keywordstyle=\bfseries,
  commentstyle=\itshape\color{black!55},
  columns=fullflexible,
  keepspaces=true,
  showstringspaces=false,
  tabsize=2,
  breaklines=true,
  breakatwhitespace=true,
  aboveskip=\smallskipamount,
  belowskip=\smallskipamount
}

\definecolor{codechange}{HTML}{1F5F99}
\lstdefinestyle{syschanges}{
  style=sys,
  basicstyle=\scriptsize\ttfamily,
  breaklines=false,
  moredelim=**[is][\color{codechange}]{@|}{|@}
}

\lstnewenvironment{syscode}[1][]{\lstset{style=sys,#1}}{}

\DeclareRobustCommand{\lean}[1]{%
  {\normalfont\scriptsize\sffamily\color{black!55}Lean:}\nobreak\hspace{0.3em}%
  {\normalfont\footnotesize\ttfamily\detokenize{#1}}%
}
\newcommand{\leanname}[1]{{\normalfont\small\ttfamily\detokenize{#1}}}
\newcommand{\leanin}[2]{\leanname{#2}}
\newcommand{\leanfile}[1]{{\scriptsize\urlstyle{tt}\def\UrlBreaks{\do\/}\def\UrlBigBreaks{}\nolinkurl{#1}}}

\newcommand{\interpchange}[1]{{\color{codechange}\boldsymbol{#1}}}

\newcommand{\ResultDSESpeedup}{1.27} 
\newcommand{\ResultDSERSSReduction}{23} 
\newcommand{\ResultTreeTwoThreeDSESpeedup}{5.21} 
\newcommand{\ResultLayoutCount}{152} 
\newcommand{\ResultBenchmarkCount}{18} 
\newcommand{\ResultMorphicOriginalCount}{15} 
\definecolor{DSEDefaultColor}{HTML}{009E73}
\definecolor{DSEFrontierColor}{HTML}{D55E00}
\DeclareRobustCommand{\DSEDefaultMarker}{%
  \tikz[baseline=-0.4ex,x=1em,y=1em]{%
    \filldraw[fill=DSEDefaultColor,draw=black,line width=0.4pt]
      (0,0.3) -- (0.3,0) -- (0,-0.3) -- (-0.3,0) -- cycle;
  }%
}
\DeclareRobustCommand{\DSEFrontierMarker}{%
  \tikz[baseline=-0.4ex,x=1em,y=1em]{%
    \draw[DSEFrontierColor,densely dotted,line width=0.8pt] (-0.7,0) -- (0.7,0);
    \filldraw[fill=DSEFrontierColor,draw=black,line width=0.35pt]
      (-0.2,-0.2) rectangle (0.2,0.2);
  }%
}
\newcommand{\ResultMLtonSpeedup}{1.53} 
\newcommand{\ResultMorphicSpeedup}{1.42} 
\newcommand{\ResultRibbitSpeedup}{1.58} 

\newcommand{\casestudyheading}[2]{%
  \subsubsection{#1}\mbox{}\par\nobreak
  {\footnotesize\noindent$#2$\par}%
  \nobreak\smallskip\noindent\ignorespaces
}

\newcommand{\KdTreeType}{%
  \RefT{KdTree}=\Ctor{Leaf}(x,y,z:\mathsf f_{32})\mid\Ctor{Node}(\mathit{min}_x,\mathit{min}_y,\mathit{min}_z,\mathit{max}_x,\mathit{max}_y,\mathit{max}_z:\mathsf f_{32},l:\RefT{KdTree},r:\RefT{KdTree})%
}
\newcommand{\TwoThreeTreeType}{%
  \RefT{Tree}=\Ctor{Leaf}\mid\Ctor{Two}(l:\RefT{Tree},k:\Uint{64},r:\RefT{Tree})\mid\Ctor{Three}(l:\RefT{Tree},k_1:\Uint{64},m:\RefT{Tree},k_2:\Uint{64},r:\RefT{Tree})%
}
\newcommand{\IntervalTreeType}{%
  \RefT{Tree}=\Ctor{Leaf}\mid\Ctor{Node}(l:\RefT{Tree},r:\RefT{Tree},\mathit{lo}:\mathsf i_{64},\mathit{hi}:\mathsf i_{64})%
}

\newcommand{\ResultKDPoints}{\ensuremath{2^{20}}} 
\newcommand{\ResultKDQueries}{\ensuremath{2^{20}}} 
\newcommand{\ResultConfidencePercent}{95} 
\newcommand{\KDMinRuntimeMarker}{\textcolor[HTML]{0072B2}{\ensuremath{\bullet}}}
\newcommand{\KDMinRSSMarker}{\textcolor[HTML]{D55E00}{\ensuremath{\bullet}}}
\newcommand{\KDParetoMarker}{\ensuremath{\circ}}

\newcommand{\ResultTreeKeys}{\ensuremath{1.6\times10^6}} 
\newcommand{\ResultTreeScanCounts}{\ensuremath{10^3\times k}} 
\newcommand{\ResultTreeScanMultipliers}{\{16,\allowbreak64,\allowbreak128,\allowbreak256,\allowbreak384,\allowbreak512\}} 
\newcommand{\ResultTreeBestFixedSpeedup}{2.29} 

\newcommand{\ResultIntervalMorphicSpeedup}{1.58} 
\newcommand{\ResultIntervalMLtonSpeedup}{1.10} 

\newcommand{\ResultWarmups}{2} 
\newcommand{\ResultSamples}{14} 
\newcommand{\ResultTrimEachTail}{2} 
\newcommand{\ResultRetainedSamples}{10} 
\newcommand{\ResultCPUCores}{10} 
\newcommand{\ResultMemoryGiB}{16} 
\newcommand{\ResultMacOS}{26.0} 
\newcommand{\ResultClang}{19.1.7} 
\newcommand{\ResultMimalloc}{3.6.1} 
\newcommand{\ResultMatchedMorphicSpeedup}{1.18} 
\newcommand{\ResultMatchedLayoutCount}{18} 
\newcommand{\ResultKDFastestLayout}{19-adopt-pad} 
\newcommand{\ResultKDSmallestLayout}{7-narrow} 
\newcommand{\ResultKDSpeedup}{1.65} 
\newcommand{\ResultKDRSSReduction}{31} 
\newcommand{\ResultTreeMaxScans}{512{,}000} 
\newcommand{\ResultIntervalQueries}{128{,}000} 
\newcommand{\ResultIntervalExponentStart}{6} 
\newcommand{\ResultIntervalExponentStop}{19} 
\newcommand{\ResultSystemMLtonSpeedup}{1.01} 
\newcommand{\ResultSystemMLtonRSSRatio}{0.29} 
\newcommand{\ResultSystemMorphicSpeedup}{1.41} 
\newcommand{\ResultSystemMorphicRSSRatio}{0.84} 
\newcommand{\ResultSystemMatchedMorphicSpeedup}{1.36} 
\newcommand{\ResultSystemRibbitOptionSpeedup}{1.45} 
\newcommand{\ResultSystemRibbitRBSpeedup}{2.22} 
\newcommand{\ResultSystemRibbitPCFSpeedup}{1.20} 
\newcommand{\ResultSystemRibbitSpeedup}{1.57} 
\newcommand{\ResultKokaCount}{4} 
\newcommand{\ResultKokaSourceCount}{5} 
\newcommand{\ResultKokaInput}{100{,}000} 
\newcommand{\ResultKokaRSSMatchedReduction}{9}
\newcommand{\ResultKokaRSSReduction}{20}
\newcommand{\ResultKoka}{3.2.9}
\newcommand{\ResultDerivImmediateRSSReduction}{50} 

\begin{document}

\title{Composable Rewrites for Layouts of Algebraic Data Types}

\author{Christophe Gyurgyik}
\orcid{0000-0001-8493-1133}
\affiliation{%
  \institution{Stanford University}
  \city{Stanford}
  \country{USA}
}
\email{cpg@cs.stanford.edu}

\author{Alexander J Root}
\orcid{0000-0001-6221-1389}
\affiliation{%
  \institution{Stanford University}
  \city{Stanford}
  \country{USA}
}
\email{ajroot@cs.stanford.edu}

\author{Benjamin Driscoll}
\orcid{0000-0002-1901-1880}
\affiliation{%
  \institution{Stanford University}
  \city{Stanford}
  \country{USA}
}
\email{bdrisc@cs.stanford.edu}

\author{Gabriel Radanne}
\orcid{0000-0002-2107-7678}
\affiliation{%
  \institution{ENS Lyon}
  \city{Lyon}
  \country{France}
}
\email{gabriel.radanne@inria.fr}

\author{Fredrik Kjolstad}
\orcid{0000-0002-2267-903X}
\affiliation{%
  \institution{Stanford University}
  \city{Stanford}
  \country{USA}
}
\email{kjolstad@cs.stanford.edu}

\author{Clément Pit-Claudel}
\orcid{0000-0002-1900-3901}
\affiliation{%
  \institution{EPFL}
  \city{Lausanne}
  \country{Switzerland}
}
\email{clement.pit-claudel@epfl.ch}

\renewcommand{\shortauthors}{Gyurgyik et al.}

\begin{abstract}
Algebraic data types (ADTs) decouple the logical structure of data from physical layout choices that affect performance. For example, packing small fields into a word improves spatial locality, while storing co-accessed fields contiguously can enable vectorization. Such choices reinforce or undermine one another, with ramifications for compiler optimizations and memory management. Their interplay creates a combinatorial space of layouts with different runtime costs and memory footprints. The most favorable layout depends on the program's construction and destruction patterns, the characteristics of input data, and the target machine.

Exploring this space requires control over layout decisions and a way to compose them while preserving program semantics. We provide both through rewrite rules over layouts, independent of any particular source language or bit-level encoding. Starting from a canonical layout derived from the ADT, these rules transform its representation while leaving the source program intact. We identify a local refinement criterion for proving each rewrite rule correct, so that layouts continue to represent every value and keep distinct values apart. Adding a new rule therefore only requires a proof satisfying this local criterion.

We realize these rules in \sys{}, a new pure functional programming language whose compiler specializes construction, pattern matching, and memory management to the chosen layouts. Using a default layout for all ADTs, \sys{} achieves competitive performance against established functional compilers that automatically select layouts. Our rewrites further improve performance by tailoring layouts to each workload, permitting representations that prior work cannot attain or can only achieve through explicit changes to the source code. Finally, \sys{} can replicate the layouts and peak memory of a language with state-of-the-art reference counting, and additional rewrites further reduce peak memory.

\end{abstract}

\maketitle

\section{Introduction}
\label{sec:introduction}
Algebraic data types (ADTs) describe data by composing products and sums: a product type holds a value of each of its component types, and a sum type holds a value of exactly one of its alternatives. This abstraction supports type-safe construction and pattern matching independently of any physical representation, and thus the same logical values can be represented using many different memory layouts. 

The choice among these layouts can substantially affect runtime performance and memory footprint, motivating extensive research on ADT layout optimizations in functional programming languages~\cite{harper1995compiling,downen2020kinds,leroy1992unboxed,shao1994unrolling,gyurgyik2026decoupling,marlow2007faster,baudon2023bitstealing,elsman2024double,peytonjones1991unboxed,chen2023dargent,eisenberg2017levity}. For example, packing small fields into a word can save space but may require additional instructions or unaligned loads when they are read, while boxing a large, rarely accessed field places it behind a pointer, keeping frequently accessed data compact at the cost of a pointer dereference for infrequently accessed data. These choices are interdependent; the benefits of boxing depend in turn on how the remaining fields are packed and ordered.

Furthermore, layouts are necessarily entangled with the language implementation. Field widths and alignment affect instruction selection and vectorization~\cite{gyurgyik2026decoupling,woop2014exploiting,benthin2018compressed,wimmer2008automatic}, while the sizes of allocated objects constrain opportunities for memory reuse~\cite{reinking2021perceus,lorenzen2023fp2,lorenzen2022reference}. A layout therefore cannot be evaluated in isolation: its performance depends on which optimizations it enables and which of those optimizations the compiler applies. More broadly, a program's construction and destruction patterns, characteristics of its input data, and the target architecture shape the Pareto frontier between runtime performance and memory footprint, leaving no universally best layout~\cite{gyurgyik2026decoupling}.

Language implementations presently address these costs through both automatic layout selection and explicit representation control. MLton and Morphic select representations through compiler analyses~\cite{weeks2006whole,brandon2023better,brandon2026fullyautomatic}. Haskell and OCaml expose representation choices through type system extensions~\cite{peytonjones1991unboxed,eisenberg2017levity,leroy2025ocaml,chataing2024unboxed}. More recently, layout description languages make the separation of data type and data layout explicit and programmable~\cite{chen2023dargent, baudon2023bitstealing, gyurgyik2026decoupling}.

We build on the decoupled approach of prior work~\cite{baudon2023bitstealing,chen2023dargent,gyurgyik2026decoupling}, introducing composable rewrite rules over a layout language for deriving ADT representations independently of a particular source language or bit-level encoding. Rather than writing each candidate layout in full, as required by layout description languages~\cite{baudon2023bitstealing,chen2023dargent,gyurgyik2026decoupling}, a programmer or an automated search procedure starts from a canonical layout induced by the source type, and transforms it through incremental rewrites. A layout is \emph{faithful} when every source value has a representation and no representation corresponds to two distinct values. Our key insight is that a local criterion is enough to preserve faithfulness: each rewrite rule need only satisfy it on the part of the layout it changes. For a core calculus, we derive construction and destruction for rewritten layouts and prove that valid rewrite sequences preserve the result of terminating programs. Adding a new rewrite rule requires only a reconstruction function that establishes the local correctness criterion.

We implement these rewrite rules in \sys{}, a pure functional programming language and compiler that embodies a separation of concerns: changing a layout never requires modifying the program. The compiler specializes construction, destruction, and memory management (reference counting) to the rewritten layouts. To support different layouts across program phases, \sys{} also provides a \code{coerce} scheduling directive, which inserts a change of representation at a selected variable binding while leaving the source code unchanged. Our primary contributions are:

\begin{itemize}[leftmargin=1.5em]
  \item Composable layout rewrite rules that, independent of a particular source language or machine, express a broad range of layouts and support extension with new rules.
  
  \item A formal model of layouts that separates structural representation from any particular bit-level encoding. Per-rule proofs compose to show that valid rewrites keep layouts faithful and preserve the scalar results of terminating programs.
  
  \item \sys{}, a new pure functional programming language that decouples layout rewrites from the source program, and a compiler that automatically specializes with respect to rewritten layouts. \sys{} additionally supports transformations of ADT layouts at arbitrary program points.
\end{itemize}

Our evaluation demonstrates that \sys{} is competitive with state-of-the-art functional languages, and our decoupled abstraction incurs no performance overhead. In three case studies, we examine composing rewrites, memoizing compute without editing the source code, and changing layouts between program phases. With an automatically selected layout, \sys{} achieves geomean speedups of $\ResultMorphicSpeedup{}\times$ over Morphic~\cite{brandon2023better} and $\ResultMLtonSpeedup{}\times$ over MLton~\cite{weeks2006whole}, whose compilers automatically select representations. We also perform layout design space exploration, where the fastest measured layout per benchmark gives a $\ResultDSESpeedup\times$ (geomean) speedup over the default, and the layout with the lowest peak memory reduces memory by \ResultDSERSSReduction\%. We compare \sys{} with Ribbit~\cite{baudon2023bitstealing}, a layout description language for general-purpose ADTs and the closest prior work. \sys{}'s code generation yields a $\ResultRibbitSpeedup\times$ (geomean) speedup when matching Ribbit's layouts. Lastly, we show layout control complements automatic memory management: with equivalent layouts, \sys{} uses at most the peak memory of Koka's fully-in-place programs~\cite{lorenzen2023fp2}, and further rewrites lower peak memory by an additional \ResultKokaRSSReduction\%.

Finally, we clarify the scope of this work. First, we assume finite trees and monomorphization in our implementation. Although monomorphization typically bars features such as polymorphic recursion and separate compilation, runtime specialization~\cite{kennedy2001design} could relax this restriction. Second, we focus on data representation, which is orthogonal to the execution strategies that user-schedulable languages decouple from algorithms~\cite{hall2025chronology, ragankelley2013halide, hagedorn2020achieving}. Combining the two in a general-purpose functional language would be a natural direction for future work. Third, our formalism covers a formal model of layouts, program preservation under layout rewrites, and recovery from physical encodings under explicit conditions. It does not establish end-to-end correctness of the \sys{} compiler.

\section{Optimizing Layouts of Red-Black Trees}
\label{sec:motivation}

\usetikzlibrary{patterns,decorations.pathreplacing}
\definecolor{motColor}{HTML}{F4C7B8}
\definecolor{motLeft}{HTML}{BED9EF}
\definecolor{motKey}{HTML}{F5DDA2}
\definecolor{motValue}{HTML}{DACAF0}
\definecolor{motRight}{HTML}{BFE2D0}
\definecolor{motMeta}{HTML}{D6DDE5}
\definecolor{motMemoized}{HTML}{F3CDE0}
\definecolor{motPad}{HTML}{F0F0F0}
\definecolor{motZero}{HTML}{FFFFFF}
\definecolor{motInk}{HTML}{425363}
\tikzset{
  mot field/.style={rectangle,draw=motInk!55,line width=.35pt,
    minimum height=.387cm,inner sep=0pt,outer sep=0pt,
    text height=1.5ex,text depth=.25ex,font=\small},
  mot width/.style={font=\scriptsize,text=motInk,inner sep=0pt,anchor=north},
  mot arrow/.style={-{Stealth[length=3.3pt,width=3.3pt]},draw=motInk,line width=.6pt}
}
\pgfmathdeclarefunction{motwidth}{1}{\pgfmathparse{.9*max(.24,.0185*#1)}}
\newcommand{\motFields}[2]{%
  \foreach \motName/\motBits/\motFill
    [count=\motIndex from 0,remember=\motEnd as \motStart (initially #1)] in {#2} {%
    \pgfmathsetmacro{\motWidth}{motwidth(\motBits)}%
    \pgfmathsetmacro{\motEnd}{\motStart+\motWidth}%
    \def\motPadding{motPad}%
    \ifx\motFill\motPadding
      \node[mot field,minimum width=\motWidth cm,
        pattern=north west lines,pattern color=motInk!40,
        anchor=west] (f\motIndex) at (\motStart,0) {};
    \else
      \node[mot field,minimum width=\motWidth cm,fill=\motFill,anchor=west]
        (f\motIndex) at (\motStart,0) {$\mathsf{\motName}$};
    \fi
    \node[mot width] at ([yshift=-2pt]f\motIndex.south) {$\motBits$};
  }%
}
\newcommand{\motRow}[1]{%
  \begin{tikzpicture}[baseline=(f0.base)]\motFields{0}{#1}\end{tikzpicture}%
}
\tikzset{
  mot icon/.style={draw=motInk!55,line width=.35pt}
}
\newcommand{\motIconBrace}{%
  \kern.12em\tikz[baseline={([yshift=-.6ex]current bounding box.center)}]{%
    \draw[motInk,line width=.9pt,decorate,decoration={brace,amplitude=1.6pt}]
      (0,-.35ex) -- (0,1.65ex);
    \draw[mot icon,fill=motColor] (.6ex,.75ex) rectangle (2.2ex,1.65ex);
    \draw[mot icon,fill=motRight] (.6ex,-.35ex) rectangle (2.2ex,.55ex);}\kern.12em%
}
\newcommand{\motIconTag}{%
  \kern.12em\tikz[baseline={([yshift=-.6ex]current bounding box.center)}]{%
    \draw[mot icon,fill=motMeta] (0,-.2ex) rectangle (1.6ex,1.5ex);
    \node[font=\tiny,inner sep=0pt] at (.8ex,.65ex) {$\mathsf{t}$};}\kern.12em%
}
\newcommand{\motIconProduct}{%
  \kern.12em\tikz[baseline={([yshift=-.6ex]current bounding box.center)}]{%
    \draw[mot icon,fill=motLeft] (0,-.25ex) rectangle (1.6ex,1.45ex);
    \draw[mot icon,fill=motKey] (1.6ex,-.25ex) rectangle (3.2ex,1.45ex);}\kern.12em%
}
\newcommand{\motIconPointer}{%
  \kern.12em\tikz[baseline={([yshift=-.6ex]current bounding box.center)}]{%
    \draw[mot icon,fill=white] (0,-.25ex) rectangle (2ex,1.45ex);
    \draw[mot arrow,line width=.5pt] (1ex,.6ex) -- (3.6ex,.6ex);
    \fill[motInk] (1ex,.6ex) circle (.9pt);}\kern.12em%
}
\newcommand{\motIconZero}{%
  \kern.12em\tikz[baseline={([yshift=-.6ex]current bounding box.center)}]{%
    \draw[mot icon,fill=white] (0,-.25ex) rectangle (1.6ex,1.45ex);
    \node[font=\tiny,inner sep=0pt] at (.8ex,.6ex) {$0$};}\kern.12em%
}
\newcommand{\motIconPad}{%
  \kern.12em\tikz[baseline={([yshift=-.6ex]current bounding box.center)}]{%
    \draw[mot icon,pattern=north west lines,pattern color=motInk!40]
      (0,-.25ex) rectangle (1.6ex,1.45ex);}\kern.12em%
}
\newcommand{\motMiniFields}[2]{%
  \foreach \motName/\motFill [count=\motIndex from 0] in {#2} {%
    \pgfmathsetmacro{\motStart}{#1+1.6*\motIndex}%
    \pgfmathsetmacro{\motMid}{\motStart+.8}%
    \def\motPadding{motPad}%
    \ifx\motFill\motPadding
      \draw[mot icon,pattern=north west lines,pattern color=motInk!40]
        (\motStart ex,-.25ex) rectangle +(1.6ex,1.7ex);
    \else
      \draw[mot icon,fill=\motFill] (\motStart ex,-.25ex) rectangle +(1.6ex,1.7ex);
      \node[font=\tiny,inner sep=0pt] at (\motMid ex,.6ex) {$\mathsf{\motName}$};
    \fi
  }%
}
\newcommand{\motMini}[1]{%
  \kern.12em\tikz[baseline={([yshift=-.6ex]current bounding box.center)}]{%
    \motMiniFields{0}{#1}}\kern.12em%
}
\newcommand{\motMiniHeap}[1]{%
  \kern.12em\tikz[baseline={([yshift=-.6ex]current bounding box.center)}]{%
    \draw[mot icon,fill=white] (0,-.25ex) rectangle (2ex,1.45ex);
    \motMiniFields{3.6}{#1}
    \draw[mot arrow,line width=.5pt] (1ex,.6ex) -- (3.6ex,.6ex);
    \fill[motInk] (1ex,.6ex) circle (.9pt);}\kern.12em%
}
\newcommand{\motStep}[1]{%
  \xrightarrow[\mathsf{#1}\vphantom{\mathsf{bp}}]{\hphantom{\mathsf{commute}}}%
}
\newcommand{\motPadSpan}[2]{%
  \begin{tikzpicture}[baseline=(f0.base)]
    \xdef\motSpanWidth{0}%
    \foreach \motBits in {#2} {%
      \pgfmathsetmacro{\motSpanNext}{\motSpanWidth+motwidth(\motBits)}%
      \xdef\motSpanWidth{\motSpanNext}%
    }%
    \node[mot field,minimum width=\motSpanWidth cm,
      pattern=north west lines,pattern color=motInk!40,
      anchor=west] (f0) at (0,0) {};
    \node[mot width] at ([yshift=-2pt]f0.south) {$#1$};
  \end{tikzpicture}%
}
\newcommand{\motHeap}[3][]{%
  \begin{tikzpicture}[baseline=(f0.base)]
    \xdef\motSpanWidth{0}%
    \ifx\relax#1\relax\else
      \motFields{0}{#1}%
      \foreach \motName/\motBits/\motFill in {#1} {%
        \pgfmathsetmacro{\motSpanNext}{\motSpanWidth+motwidth(\motBits)}%
        \xdef\motSpanWidth{\motSpanNext}%
      }%
    \fi
    \pgfmathsetmacro{\motPointerWidth}{motwidth(#2)}%
    \pgfmathsetmacro{\motPayloadStart}{\motSpanWidth+\motPointerWidth+.45}%
    \node[mot field,minimum width=\motPointerWidth cm,fill=white,anchor=west]
      (p) at (\motSpanWidth,0) {};
    \node[mot width] at ([yshift=-2pt]p.south) {$#2$};
    \motFields{\motPayloadStart}{#3}
    \draw[mot arrow] (p.center) -- (f0.west);
    \fill[motInk] (p.center) circle (1.35pt);
  \end{tikzpicture}%
}

Consider a programmer implementing an ordered set as a red-black tree~\cite{okasaki1999red}:
\[
\RefT{Color}=\Ctor{Red}\mid\Ctor{Black}\qquad
\RefT{Tree}=\Ctor{Leaf}\mid\Ctor{Node}(\mathsf c:\RefT{Color},\,\mathsf l:\RefT{Tree},\,\mathsf k:\Uint{64},\,\mathsf r:\RefT{Tree})
\]
Throughout this section, the type and source code stay fixed while the physical layout changes. The layout of $\RefT{Tree}$ is defined recursively and each child field, $\mathsf l$ and $\mathsf r$, uses the same layout. We assume little-endian byte order, 64-bit pointers and 8-byte aligned allocations. Our diagrams place storage offsets in increasing order from left to right, with each field's least significant bit (LSB) on the left. Adjacency~(\motIconProduct) represents contiguous memory, braces~(\motIconTag\motIconBrace) distinguish arms discriminated by a value $t$, arrows~(\motIconPointer) lead from pointers to heap cells, and zero bits~(\motIconZero) are marked separately from padding~(\motIconPad), whose contents are undefined.

A \emph{rewrite} makes one elementary change to a layout, e.g., inserting unused bits before a field ($\textsf{pad})$, moving a field into a heap cell ($\textsf{box}$), or swapping two adjacent fields ($\textsf{commute})$:
\[
\motMini{c/motColor}\motStep{pad}\motMini{pad/motPad,c/motColor}
\hspace{5em}
\motMini{k/motKey}\motStep{box}\motMiniHeap{k/motKey}
\hspace{5em}
\motMini{c/motColor,l/motLeft}\motStep{commute}\motMini{l/motLeft,c/motColor}
\]
Each layout below is realized with a sequence of such rewrites. We start with the layout that Morphic~\cite{brandon2023better,brandon2026fullyautomatic} automatically selects for this type and explore alternative layouts. This initial layout follows a C-style tagged union with natural alignment, in which a tree value is a byte tag (or discriminant) beside a pointer to a separately allocated heap cell. Fields are aligned, and those that are less than one word are padded to a full word (8 bytes). A tree value therefore occupies 16 bytes, and the node's fields occupy 48 bytes, including both 16-byte children.
\begin{listingdisplay}{lst:morphic}
\motRow{t/1/motMeta,0/7/motZero}\;
\left\{\begin{array}{@{\hspace{3pt}}ll@{}}
  \motPadSpan{120}{56,64}
    &\text{if }t\in\Ctor{Leaf},\\[12pt]
  \motHeap[pad/56/motPad]{64}{c/1/motColor,0/7/motZero,pad/56/motPad,l/128/motLeft,k/64/motKey,r/128/motRight}
    &\text{if }t\in\Ctor{Node}.
\end{array}\right.
\end{listingdisplay}
While this layout permits aligned memory loads, nearly half of the bits in the $\Ctor{Node}$'s fields are zero or padding. Systems programmers reclaim such bits by hand. The Linux kernel's implementation of a red-black tree, for example, stores a $\Ctor{Node}$'s color in an unused low bit of its parent pointer~\cite{torvalds2026rbtree}. This decision also exposes clever instruction selection. Red is encoded as zero, so a red $\Ctor{Node}$'s parent field stores exactly its parent's address, and tree insertion can dereference it without masking~\cite{lespinasse2012rbtree}. Similarly, a pointer to a black node can be effectively encoded as $p + 1$ where $p$ is the pointer address, resulting in a single instruction, \code{lea}, on x86~\cite{goldstein2023rbtree}. The same opportunity exists in our tree: moving each child's tag into a low bit of its pointer shrinks each to one word. We also hoist the color of each $\Ctor{Node}$ into a second low bit of the pointer that references it, resulting in a 24-byte cell:
\begin{listingdisplay}{lst:low-bits}
\motRow{t/1/motMeta}\;
\left\{\begin{array}{@{\hspace{3pt}}ll@{}}
  \motPadSpan{63}{1,62}&\text{if }t\in\Ctor{Leaf},\\[12pt]
  \motHeap[c/1/motColor]{62}{l/64/motLeft,k/64/motKey,r/64/motRight}
    &\text{if }t\in\Ctor{Node}.
\end{array}\right.
\end{listingdisplay}
The layout applies recursively to both children: each occupies one word, consisting of a one-bit tag followed by either a color bit and a 62-bit pointer for a $\Ctor{Node}$ or 63 bits of padding for a $\Ctor{Leaf}$. An additional benefit is that red-black tree rebalancing can test children's colors directly without first dereferencing a pointer.

But we can go further: by default, x86-64 Linux places user allocations below $2^{47}$~\cite{linux2026fivelevel}, leaving the upper 17 bits of their 64-bit addresses zero. An 8-byte-aligned address thus needs only bits $[3,47)$. This enables a \emph{double-ended} bit stealing technique~\cite{elsman2024double}, where a pointer is stored in just 44 bits. Consequently, a child now occupies only 47 bits of space, and some additional padding keeps each cell's fields at byte boundaries:
\begin{listingdisplay}{lst:double-ended}
\motRow{t/1/motMeta}\;
\left\{\begin{array}{@{\hspace{3pt}}ll@{}}
  \motPadSpan{46}{1,1,44}&\text{if }t\in\Ctor{Leaf},\\[12pt]
  \motHeap[c/1/motColor,pad/1/motPad]{44}{l/47/motLeft,pad/1/motPad,r/47/motRight,pad/1/motPad,k/64/motKey}
    &\text{if }t\in\Ctor{Node}.
\end{array}\right.
\end{listingdisplay}

So far, these rewrites reclaim unused bits in an effort to reduce the memory footprint. The tag can \emph{also} distinguish more interesting tree shapes, which lets a layout store such shapes more compactly. Consider a $\Ctor{Node}$ whose children are both $\Ctor{Leaf}$, such as the one each insertion creates for its new key. Layout~\ref{lst:double-ended} still reserves $47+47=94$ bits for these two children, even though they provide no additional information. We can instead give this shape an arm of its own, $\Ctor{Node}(\Ctor{Leaf},\Ctor{Leaf})$, similar to prior work in acceleration structure design~\cite{benthin2018compressed}. Within this arm, an $\textsf{unfold}$ rewrite expands each child's layout, and, since the layout of $\Ctor{Leaf}$ is empty, $\textsf{l}$ and $\textsf{r}$ do not need to be stored. With the color already in the reference, this arm's cell then holds only its key, in exactly 8 bytes. With three arms, the tag widens to two bits, and every other $\Ctor{Node}(\mathord{\vcenter{\hbox{$\scriptscriptstyle\bullet$}}},\mathord{\vcenter{\hbox{$\scriptscriptstyle\bullet$}}})$ shape falls back to the general $\Ctor{Node}$ layout. Its cell is unchanged, and the wider tag takes the place of the padding bit before its pointer:
\begin{listingdisplay}{lst:shape}
\motRow{t/2/motMeta}\;
\left\{\begin{array}{@{\hspace{3pt}}ll@{}}
  \motPadSpan{45}{1,44}
    &\text{if }t\in\Ctor{Leaf},\\[12pt]
  \motHeap[c/1/motColor]{44}{k/64/motKey}
    &\text{if }t\in\Ctor{Node}(\Ctor{Leaf},\Ctor{Leaf}),\\[12pt]
  \motHeap[c/1/motColor]{44}{l/47/motLeft,pad/1/motPad,r/47/motRight,pad/1/motPad,k/64/motKey}
    &\text{if }t\in\Ctor{Node}.
\end{array}\right.
\end{listingdisplay}

Further rewrites can exploit properties of the data. Keys may be drawn from a much smaller distribution than the type admits. A $\textsf{refine}$ rewrite gives the specialized $\Ctor{Node}$ whose keys are below $2^{32}$ an arm of their own, which stores the key in 32 bits. As the color already sits in the reference, an $\textsf{unbox}$ rewrite can store the key there too, in place of the pointer, leaving this arm with no cell at all. The layout maintains the full \code{u64} domain, but incurs a cost of checking whether the key fits in 32 bits when constructing a node:
\begingroup
\pgfmathsetmacro{\motRefWidth}{
  motwidth(1)+motwidth(5)+motwidth(32)+motwidth(7)
}
\newcommand{\motTerminalHeap}[1]{%
  \begin{tikzpicture}[baseline=(f0.base)]
    \pgfmathsetmacro{\motColorWidth}{motwidth(1)}%
    \pgfmathsetmacro{\motPointerWidth}{\motRefWidth-\motColorWidth}%
    \pgfmathsetmacro{\motPayloadStart}{\motRefWidth+.45}%
    \motFields{0}{c/1/motColor}
    \node[mot field,minimum width=\motPointerWidth cm,
      fill=white,anchor=west] (p) at (\motColorWidth,0) {};
    \node[mot width] at ([yshift=-2pt]p.south) {$44$};
    \motFields{\motPayloadStart}{#1}
    \draw[mot arrow] (p.center) -- (f0.west);
    \fill[motInk] (p.center) circle (1.35pt);
  \end{tikzpicture}%
}

\begin{listingdisplay}{lst:terminal}
\motRow{t/2/motMeta}\;
\left\{\begin{array}{@{\hspace{3pt}}ll@{}}
  \motPadSpan{45}{1,5,32,7}
    &\text{if }t\in\Ctor{Leaf},\\[12pt]
  \motRow{c/1/motColor,pad/5/motPad,k/32/motKey,pad/7/motPad}
    &\text{if }t\in\Ctor{Node}(\Ctor{Leaf},\Ctor{Leaf})
      \land\mathsf k<2^{32},\\[12pt]
  \motTerminalHeap{k/64/motKey}
    &\text{if }t\in\Ctor{Node}(\Ctor{Leaf},\Ctor{Leaf})
      \land\mathsf k\ge2^{32},\\[12pt]
  \motTerminalHeap{
    l/47/motLeft,pad/1/motPad,
    r/47/motRight,pad/1/motPad,k/64/motKey}
    &\text{if }t\in\Ctor{Node}.
\end{array}\right.
\end{listingdisplay}
\endgroup

Finally, layouts also interact with memory management. Rebalancing a tree consumes and rebuilds nodes, and the compiler can reuse the allocation of a consumed cell for a rebuilt node \emph{if} they share the same size~\cite{ullrich2019counting,reinking2021perceus}. In Layout~\ref{lst:terminal}, a $\Ctor{Node}(\Ctor{Leaf},\Ctor{Leaf})$ with a 64-bit key outgrows its 8-byte cell when an insertion replaces one of its $\Ctor{Leaf}$ children, forcing a fresh allocation. A $\textsf{pad}$ rewrite restores reuse by growing these cells to 20 bytes, which prior work on reference counting achieves by modifying the source program itself~\cite{lorenzen2023fp2}.

We express optimizations beyond this example, such as array inlining~\cite{wimmer2008automatic} and memoizing values that are expensive to recompute~\cite{openjdk2025hashmap}. Each layout exploits a semantics-preserving transformation of the program, and the best combination depends on the workload in ways that fixed compiler heuristics cannot easily realize. A functional programmer \emph{today} can express only a handful of these layouts, and even those often require manipulating types or the source program itself~\cite{peytonjones1991unboxed,lorenzen2024oxcaml,lorenzen2023fp2}.

\section{Overview}
\label{sec:overview}

We now present a general framework for expressing and composing such layout transformations. \Cref{sec:layout-language} introduces a layout language that can describe each layout above, together with rewrite rules that each make a local change, e.g., swapping two fields or placing a field behind a pointer. Layout rewriting starts from a source type's canonical layout, which mirrors its type declaration.

\Crefrange{sec:representation-semantics}{sec:physical-storage} develop the formalism. \Cref{sec:representation-semantics} defines the \emph{structural representations} a layout permits for each source value, which describe how the layout organizes the value before it is encoded as bits. \Cref{sec:rewrite-correctness} proves each rule correct on the part of the layout it changes and lifts this guarantee through enclosing and recursive layouts, so every layout reached from the canonical one by valid rewrites is faithful, and terminating programs in a core calculus compute the same scalar results. \Cref{sec:physical-storage} gives conditions under which a model encoder and decoder recover source values from memory. The development is mechanized in Lean\footnote{The mechanization relies upon two axioms, propositional extensionality and quotient soundness.}~\cite{moura2021lean}, and references identify Lean declarations corresponding to formal statements and grammars, e.g., \lean{WF}. \Aref{app:mechanization} indexes the complete set of definitions, theorems, and worked examples.

\Cref{sec:implementation} describes the implementation, \sys{}, a new functional language that uses our rewrites to derive data layouts. Its compiler specializes construction, destruction, and reference counting to the resulting layouts, and its \code{coerce} directive changes a value's layout between program phases that favor different layouts. Finally, we evaluate the runtime performance and memory footprint of \sys{} programs in \Cref{sec:evaluation}, through case studies and comparisons with state-of-the-art functional languages.

\begingroup\setlength{\emergencystretch}{2em} 
\section{Layout Language}
\label{sec:layout-language}

A \emph{layout} describes how the values of a source type are represented (grammar in \Cref{fig:layout-grammar}). Before defining the individual constructs, we express Layout~\ref{lst:double-ended} from \Cref{sec:motivation} in the layout language:
{\small
\begin{equation*}
\begin{aligned}
t:\Uint1\times\Tagged t{&\ar{(0,\Ctor{Leaf})}{\Pad{46}},\\
&\ar{(1,\Ctor{Node})}{\mathsf c:\RefL{Color}\times\Pad1\times\Ptr{44}{\mathsf l:\RefL{Tree}\times\Pad1\times\mathsf r:\RefL{Tree}\times\Pad1\times\mathsf k:\Uint{64}}}}
\end{aligned}
\end{equation*}
}%
Read from left to right, a tree stores a one-bit tag $t$, followed by the arm that $t$ selects, code $0$ for a $\Ctor{Leaf}$ and $1$ for a $\Ctor{Node}$. A $\Ctor{Leaf}$ stores 46 bits of padding, while a $\Ctor{Node}$ stores its color, a bit of padding, and a 44-bit pointer to its remaining fields. The color $\textsf{c}$ refers to the layout of $\RefT{Color}$, which stores only a one-bit tag. Each child refers back to the layout of $\RefT{Tree}$, thus each occupies $1+1+1+44=47$ bits. The rest of this section defines these constructs and the rewrite rules that derive this layout from the one induced by the source type.

\subsection{Values and Paths}
\label{sec:layout-paths}
Source values are unsigned scalars and finite constructor trees $\Ctor K(v_1,\dots,v_n)$, with fields in the order originally given by the source type. We write $\mathcal V$ for all source values and $\mathcal V_\tau\subseteq\mathcal V$ for all source values of an algebraic data type $\tau$. We call the source value that a layout represents its \emph{subject}. A \emph{path} $\pi$ selects a subvalue, or a range of bits of a scalar, and write $v.\pi$ for the value that $\pi$ selects from $v$, when it exists. The empty path $\varepsilon$ selects $v$ itself, the path $x$ selects its field $x$, and $x[i,j)$ selects bits $[i,j)$ of that field, counting from the least significant bit. For example, in node $v=\Ctor{Node}(\Ctor{Black},\Ctor{Leaf},5,\Ctor{Leaf})\in\mathcal V_{\RefT{Tree}}$, the path $\mathsf k$ selects the key, $v.\mathsf k=5$, and $\mathsf k[0,2)$ selects its two low bits, $v.\mathsf k[0,2)=1$. For simplicity, we assume that a field name shared by constructors always has the same type.

\subsection{Provenances}
\label{sec:layout-provenances}

A layout can exploit knowledge of a value's constructors to specialize its representation~\cite{shao1994unrolling,benthin2018compressed}. Following~\citet{baudon2023bitstealing}, we call this knowledge a \emph{provenance}. We depart from their patterns by taking each provenance to be the set of values it denotes, so that provenances form a lattice. The provenance $\pi\mapsto\Ctor K$ denotes the source values whose subvalue at path $\pi$ has constructor $\Ctor K$, and we abbreviate $\varepsilon\mapsto\Ctor K$ as $\Ctor K$. Since a path can be arbitrarily long, a provenance can test a subvalue at any depth, such as $\mathsf l.\mathsf r\mapsto\Ctor{Leaf}$ for a node whose left child's right child is a $\Ctor{Leaf}$. A scalar guard $\pi\mapsto c$ similarly selects the values with scalar $c$ at path $\pi$, e.g., Layout~\ref{lst:terminal} guards its small keys by $\bigcup_{i=0}^{2^{32}-1}\mathsf k\mapsto i$. As sets of values, provenances form a lattice bounded by $\top=\mathcal V$ and $\bot=\varnothing$, with intersection as meet, union as join, and inclusion as order. Two provenances are equal when they denote the same values. \Aref{app:provenance-lattice} provides the complete depth-one lattice of the red-black $\RefT{Tree}$. Such provenances readily model patterns on ADTs, e.g.,
\[ 
\Ctor{Node}(\_,\Ctor{Node}(\Ctor{Black},\_,\_,\_),\_,\_) \;\equiv\; (\varepsilon \mapsto \Ctor{Node} \cap l \mapsto \Ctor{Node} \cap l.c \mapsto \Ctor{Black}).
\]
Each layout must represent a set $A\subseteq\mathcal V$ of \emph{admitted subjects}, the source values that can reach it. An arm has a condition $\phi$, its \emph{guard}, and a layout for the values in $A\cap\phi$, its \emph{payload}. A sum offers several arms, whose guards must together cover $A$. We additionally allow overlapping guards as an explicit design decision, since some optimizations rely on this~\cite{benthin2018compressed,hall1994using}.

\subsection{Layouts}

\begin{figure}[t]
\centering
\footnotesize
\renewcommand{\arraystretch}{0.9}
\(
\begin{array}{@{}r@{\;}c@{\;}l@{\quad\;}l@{\hspace{3em}}r@{\;}c@{\;}l@{\quad}l@{}}
  \multicolumn{8}{c}{x,\RefT T,\Ctor K\in\mathsf{Id},\qquad n,i,j,c\in\mathbb N}\\[0.6ex]
  \ell&::=&\epsilon\mid\pi:\ell&\text{empty, name}
   &\pi&::=&\varepsilon\mid x.\pi\mid\pi[i,j)&\text{names and bit subsets}\\
  &&\mid\Pad n\mid\Zero n\mid\Uint n\mid\Stored e&\text{leaves}
   &e&::=&n\mid\pi\mid e\oplus e&\text{memoized expressions}\\[0.4ex]
  &&\mid\Ptr n\ell&\text{pointer}
   &\phi&::=&\top\mid\bot&\text{provenance}\\
  &&\mid\RefL T&\text{recursive reference}
   &&&\mid\pi\mapsto\Ctor K\mid\pi\mapsto c&\text{constructor, scalar}\\
  &&\mid\ell_0\times\ell_1&\text{product}
   &&&\mid\phi\cap\phi\mid\phi\cup\phi&\text{intersection, union}\\[0.4ex]
  &&\mid\Sum{\ar{\phi_j}{\ell_j}}&\text{sum}
   &\Delta&::=&\varnothing\mid\Delta,\,\RefT T\mapsto\ell&\text{layout environment}\\
  &&\mid\Tagged{\pi}{\ar{(c_j, \phi_j)}{\ell_j}}&\text{disjoint sum} &&&&
\end{array}
\)

\caption{Grammar of the layout language. \lean{Layout, Path, Expr, Prov, LayoutEnv}}
\label{fig:layout-grammar}
\end{figure}

To let rewrites target individual layout choices, the language makes explicit how a value is stored. A named layout $\pi:\ell$ makes $v.\pi$ the subject of $\ell$, and $\epsilon$ is the empty layout, which stores nothing. Leaves describe scalar storage: $\Pad n$ reserves $n$ bits with unconstrained contents, and $\Zero n$ requires them to be zero. Layout~\ref{lst:morphic}, for instance, follows its one-bit tag with $\Zero7$ to fill a byte. $\Uint n$ stores its scalar subject in $n$ bits. A \emph{memoized expression} $\Stored e$ stores the result of evaluating $e$ on the subject, at the width of its resulting type. Expressions use integer literals, scalar path reads, and supplied binary operations, but can't use arbitrary layout components. A pointer $\Ptr n\ell$ places the representation described by $\ell$ behind an $n$-bit address. A \emph{layout environment} $\Delta$ maps type names to layouts. An entry $\Delta(\RefT T)=\ell$ is a \emph{layout definition}, with body $\ell$, and a reference $\RefL T$ stands for that body.

Products $\ell_0\times\ell_1$ are binary and ordered; we omit the parentheses of right-associated products. Their components occupy increasing bit offsets from left to right, and offsets restart at zero in each pointee. The cell of Layout~\ref{lst:double-ended} places $\mathsf l$ at offset 0, $\mathsf r$ at 48, and $\mathsf k$ at 96. 
An ordinary sum $\Sum{\ar{\phi_j}{\ell_j}}$ has arms with distinct guards and represents a subject by the payload of any arm whose guard it satisfies; we name each arm by its guard. A \emph{disjoint sum} $\Tagged\pi{\ar{(c_j,\phi_j)}{\ell_j}}$ instead selects its arm by the scalar at path $\pi$, using arm $j$ for every subject whose scalar at $\pi$ equals the code $c_j$. Its codes are distinct, and each guard $\phi_j$ must contain every admitted subject with code $c_j$. 

The double-ended layout of Layout~\ref{lst:double-ended} starts as an ordinary sum over $\Ctor{Leaf}$ and $\Ctor{Node}$, and tagging turns it into a disjoint sum keyed by its tag field $t$ (\Cref{fig:rewrite-example}). A \emph{tag} under path $t$ names the scalar function that computes each value's code. The tag lists $k$ guards, $\mathsf{guards}(t)$, and a selector $s_t$ that maps each value $v$ in their union to the index $j$ of a guard $\phi_j$ containing $v$. The path $t$ selects this index, $v.t=s_t(v)$, and is undefined on every other value. The selector resolves overlapping guards, so their order imposes no priority. A tag field $t:\Uint{|t|}$ stores the code in  $|t| = \lceil \log_2 k \rceil$ bits.

Finally, source types induce a straightforward \emph{canonical} layout environment $\Delta_{\mathrm{can}}$. It binds each declared type to a sum with one arm per constructor $\Ctor K$, whose payload is the product of the constructor's named fields, or $\epsilon$ if there are none, and a field of a declared type refers to that type. For the original red-black $\RefT{Tree}$, we write
\[
\Delta_{\mathrm{can}}(\RefT{Tree})=\Sum{\ar{\Ctor{Leaf}}{\epsilon},\quad\ar{\Ctor{Node}}{\mathsf c:\RefL{Color}\times\mathsf l:\RefL{Tree}\times\mathsf k:\Uint{64}\times\mathsf r:\RefL{Tree}}}.
\]

\subsection{Layout Rewrites}
\label{sec:layout-rewrites}

\begin{figure}[t]
\centering
\footnotesize
\begin{minipage}[t]{0.44\linewidth}
\vspace{0pt}
\centering
\(\begin{array}{@{}r@{\;}c@{\;}l@{}}
\ell&\xrightarrow{\mathsf{pad}_n}&\Pad n\times\ell\\[0.9ex]
\ell&\xrightarrow{\mathsf{zero}_n}&\Zero n\times\ell\\[0.9ex]
\ell&\xrightarrow{\mathsf{ptr}_n}&\Ptr n\epsilon\times\ell\\[0.9ex]
\Ptr n{\ell_0}\times\ell_1&\xrightarrow{\mathsf{box}}&
    \Ptr n{\ell_0\times\ell_1}\\[0.9ex]
\Ptr n\ell&\xrightarrow{\mathsf{unbox}}&\ell\\[0.9ex]
\ell_0\times\ell_1&\xrightarrow{\mathsf{commute}}&\ell_1\times\ell_0\\[0.9ex]
(\ell_0\times\ell_1)\times\ell_2&\xleftrightarrow{\mathsf{assoc}}&\ell_0\times(\ell_1\times\ell_2)\\[0.9ex]
\epsilon\times\ell&\xleftrightarrow{\mathsf{unit}}&\ell\\[0.9ex]
\pi:\Uint n&\xrightarrow{\mathsf{split}_k}&
    \pi[0,k):\Uint k\times\pi[k,n):\Uint{n-k}
\end{array}\)
\end{minipage}\hfill
\begin{minipage}[t]{0.56\linewidth}
\vspace{0pt}
\centering
\(\begin{array}{@{}r@{\;}c@{\;}l@{}}
\Sum{\ar{\phi_j}{\ell_j}}&\xrightarrow{\mathsf{tag}}&
    t:\Uint n\times\Tagged t{\ar{(j,\phi_j)}{\ell_j}}\\[0.9ex]
\ell_0\times\Sum{\ar{\phi_j}{\ell_j}}&\xrightarrow{\mathsf{distribute}}&
    \Sum{\ar{\phi_j}{\ell_0\times\ell_j}}\\[0.9ex]
\Sum{\ar{\phi_j}{\ell\times\ell_j}}&\xrightarrow{\mathsf{factor}}&
    \ell\times\Sum{\ar{\phi_j}{\ell_j}}\\[0.9ex]
\ell&\xrightarrow{\mathsf{refine}_\phi}&
    \Sum{\ar\phi\ell,\ar\top\ell}\\[0.9ex]
\Sum{\ar\phi\ell,\Omega}&\xrightarrow{\mathsf{prune}}&
    \Sum\Omega\\[0.9ex]
\Sum{\ar\phi\ell}&\xrightarrow{\mathsf{collapse}}&\ell\\[0.9ex]
\Sum{\ar\phi{\Sum{\ar{\psi_j}{\ell_j}}}}&\xrightarrow{\mathsf{flatten}}&
    \Sum{\ar{\phi\cap\psi_j}{\ell_j}}\\[0.9ex]
\RefL T&\xrightarrow{\mathsf{unfold}}&\Delta(\RefT T)\\[0.9ex]
\ell&\xrightarrow{\mathsf{memoize}_{e}}&\Stored e\times\ell
\end{array}\)
\end{minipage}
\caption{Layout rewrites, with premises omitted. See \Aref{app:rewrite-judgments} for rules with premises (\lean{Core.Step}).}
\label{fig:rewrite-rules}
\end{figure}

\Cref{fig:rewrite-rules} lists a representative subset of rewrite rules used in our implementation, each of which makes one local change to a layout. Commuting fields places data that is read together side by side, and boxing moves rarely read fields out of line behind a pointer. Tagging stores the choice of arm in a tag field that subsequent rewrites can transform like any other field, and memoizing an expression stores a computed result within the layout so that a program reads the result in place of recomputing it. Each rule carries premises, which \Aref{app:rewrite-judgments} states in full. For example, splitting an $n$-bit field at $k$ requires $k\le n$, and pruning an arm requires the remaining arms to cover every admitted subject. Crucially, the rule set is extensible because its proofs are modular. As~\Cref{sec:extending-rewrites} shows, a new rewrite rule only needs a proof for the part of the layout it changes.

A rewrite applies a rule to one occurrence of a sublayout, its \emph{focus}, and leaves the rest of the layout fixed. A \emph{layout context} $\Ctx[\square]$ is the surrounding layout with the focus replaced by a hole, and filling the hole with $\ell$ gives $\Ctx[\ell]$. The grammar is
\[
\Ctx::=\square\mid\pi:\Ctx\mid\Ptr n \Ctx\mid \Ctx\times\ell\mid\ell\times \Ctx\mid\Sum{\ar{\phi}{\Ctx},\Omega}\mid\Tagged\pi{\ar{(c,\phi)}{\Ctx},\Omega}.
\]
For example, $\Ctx[\square]=\mathsf k:\Uint{64}\times\square$ keeps the field $\mathsf{k}$ fixed while rewriting the component in the hole. A rule that replaces $\ell_0$ by $\ell_1$ takes $\Ctx[\ell_0]$ to $\Ctx[\ell_1]$, and its premises are checked on the subjects that reach the hole through the context's field selections and guards. We write $\ell_0\longrightarrow^*\ell_k$ for a finite sequence of rewrites.

Rewriting the body of a type updates the layout environment. An \emph{update} $\Delta'=\Delta[\RefT T\mapsto\ell_1]$ replaces the body assigned to $\RefT T$ and leaves other entries unchanged, so every reference $\RefL T$, including recursive references, then uses $\ell_1$. \Cref{fig:rewrite-example} derives Layout~\ref{lst:double-ended} this way, rewriting the canonical body of $\RefT{Tree}$ one focus at a time, and \Cref{sec:context-replacement} gives the conditions under which replacements in contexts and updates to definitions preserve faithfulness.

\begin{figure}[!t]
\centering
\scriptsize
\setlength{\jot}{0pt}
\setlength{\belowdisplayskip}{0pt}
\setlength{\belowdisplayshortskip}{0pt}
\[
\begin{aligned}
&\Sum{\ar{\Ctor{Leaf}}{\interpchange{\epsilon}},\quad
  \ar{\Ctor{Node}}{\mathsf c\times\mathsf l\times\mathsf k\times\mathsf r}}
\\
\xrightarrow{\mathsf{pad}_{46},\ \mathsf{commute},\ \mathsf{unit}}\quad&
\Sum{\ar{\Ctor{Leaf}}{\Pad{46}},\quad
  \ar{\Ctor{Node}}{\mathsf c\times\mathsf l\times\interpchange{\mathsf k\times\mathsf r}}}
\\
\xrightarrow{\mathsf{pad}_{1},\ \mathsf{commute},\ \mathsf{pad}_{1}}\quad&
\Sum{\ar{\Ctor{Leaf}}{\Pad{46}},\quad
  \ar{\Ctor{Node}}{\mathsf c\times\interpchange{\mathsf l\times\Pad1\times\mathsf r\times\Pad1\times\mathsf k}}}
\\
\xrightarrow{\mathsf{ptr}_{44},\ \mathsf{box},\ \mathsf{unit}}\quad&
\Sum{\ar{\Ctor{Leaf}}{\Pad{46}},\quad
  \ar{\Ctor{Node}}{\mathsf c\times\interpchange{\Ptr{44}{\mathsf l\times\Pad1\times\mathsf r\times\Pad1\times\mathsf k}}}}
\\
\xrightarrow{\mathsf{pad}_{1}}\quad&
\interpchange{\Sum{\ar{\Ctor{Leaf}}{\Pad{46}},\quad
  \ar{\Ctor{Node}}{\mathsf c\times\Pad1\times\Ptr{44}{\mathsf l\times\Pad1\times\mathsf r\times\Pad1\times\mathsf k}}}}
\\
\xrightarrow{\mathsf{tag}}\quad&
t:\Uint1\times\Tagged t{\ar{(0,\Ctor{Leaf})}{\Pad{46}},\quad
  \ar{(1,\Ctor{Node})}{\mathsf c\times\Pad1\times\Ptr{44}{\mathsf l\times\Pad1\times\mathsf r\times\Pad1\times\mathsf k}}}
\end{aligned}
\]
\caption{Rewrites from the canonical body of $\RefT{Tree}$ to Layout~\ref{lst:double-ended}, abbreviating each field by its name. Each arrow rewrites the preceding layout at its $\interpchange{\textbf{bolded}}$ focus.}
\label{fig:rewrite-example}
\end{figure}

\section{Structural Representation Semantics}
\label{sec:representation-semantics}

To reason about rewrites locally without tying correctness to a particular bit-level encoding, we first reason about \emph{structural representations}: finite trees that retain the layout's components and their organization. An encoder then places these components in bits and heap blocks, and a decoder recovers them. Because a structural representation keeps boundaries between components explicit, we show rewrites can be proved correct without reasoning about memory in~\Cref{sec:rewrite-correctness}. In~\Cref{sec:physical-storage}, we encode these representations in bits, where boundaries are implicit, and discuss limitations.

\subsection{Structural Representations}
\label{sec:structural-values}

Structural representations $\eta\in\IR$ are defined by the grammar
\[
\eta::=k\mid()\mid(\eta_0,\eta_1)\mid\cons{\phi}\eta\mid\PtrVal\alpha\eta. \;\;\text{\lean{Enc}}
\]
Scalars $k \in \mathbb N$ hold source data, padding, tags or the results of memoized expressions, at widths supplied by the layout. A product contributes a pair, and a pointer records an address $\alpha$ together with its pointee's representation. An ordinary sum contributes a marker $\cons{\phi}\eta$ naming the selected arm by its guard. The marker has no physical counterpart; it keeps arms distinguishable until a tag rewrite stores the choice in a field. A disjoint sum contributes the pair $(c,\eta)$ of its selected code and payload; physical encoding stores the code only in the tag field.

For a layout $\ell$ under environment $\Delta$, $\W^{\Delta}_{\ell}(v)$ is the set of structural representations of subject $v$ (defined in~\Cref{fig:layout-interpretation}). Sets accommodate unspecified padding and addresses, as well as overlapping arms. The canonical layout gives a node exactly one structural representation, e.g.,
\[
  \W\bigl(\Ctor{Node}(\Ctor{Black},\Ctor{Leaf},5,\Ctor{Leaf})\bigr)=\bigl\{\cons{\Ctor{Node}}{\bigl(\cons{\Ctor{Black}}{\smallunit},\ \cons{\Ctor{Leaf}}{\smallunit},\ 5,\ \cons{\Ctor{Leaf}}{\smallunit}\bigr)}\bigr\}.
\]

\begin{figure}[b]
\centering
\begingroup\footnotesize
\setlength{\jot}{0pt}
$\begin{alignedat}{2}
\W^{\Delta}_\epsilon(v)&=\{()\},\quad \W^{\Delta}_{\Uint n}(k)=\{k\}&\qquad \W^{\Delta}_{\pi:\ell}(v)&=\W^{\Delta}_\ell(v.\pi)\\
\W^{\Delta}_{\Pad n}(v)&=\mathcal V_{\Uint n}&\qquad \W^{\Delta}_{\ell_0\times\ell_1}(v)&=\W^{\Delta}_{\ell_0}(v)\times\W^{\Delta}_{\ell_1}(v)\\
\W^{\Delta}_{\Zero n}(v)&=\{0\}&\qquad \W^{\Delta}_{\Ptr n\ell}(v)&=\{\PtrVal{\alpha}{\eta}\mid \alpha\in\mathcal V_{\Uint n},\ \eta\in\W^{\Delta}_\ell(v)\}\\
\W^{\Delta}_{\Stored e}(v)&=\{\llbracket e\rrbracket(v)\}&\qquad \W^{\Delta}_{\Sum{\ar{\phi_j}{\ell_j}}}(v)&=\{\cons{\phi_j}\eta\mid v\in\phi_j,\ \eta\in\W^{\Delta}_{\ell_j}(v)\}\\
\W^{\Delta}_{\RefL T}(v)&=\W^{\Delta}_{\Delta(\RefT T)}(v)&\qquad \W^{\Delta}_{\Tagged\pi{\ar{(c_i,\phi_i)}{\ell_i}}}(v)&=\{(c_i,\eta)\mid v.\pi=c_i,\ \eta\in\W^{\Delta}_{\ell_i}(v)\}
\end{alignedat}$
\endgroup
\par\vspace{3pt}
\caption{Structural interpretation of layouts. $\W^{\Delta}_\ell(v)$ is the set of structural representations permitted by layout $\ell$ for source value $v$. The layout environment $\Delta$ maps declared types to layout bodies. We write $\mathcal{V}_{\Uint n}$ to represent the set $\{0, \dots, 2^n - 1\}$, and $\W^{\Delta}_\ell(v)=\varnothing$ when evaluation is undefined. \lean{Rep}}
\label{fig:layout-interpretation}
\end{figure}

\subsection{Faithfulness}
\label{sec:faithfulness}

To preserve programs over different layouts, a layout must be \emph{total}, representing every source value, and \emph{absolutely injective}, giving distinct values disjoint sets of representations. Totality ensures that a valid construction policy exists, one that chooses a permitted representation for every value a program builds. Absolute injectivity ensures that destruction has a unique answer.

\begin{definition}[Faithfulness \lean{Faithful}]
\label[definition]{def:faithful}
A layout $\ell$ \emph{faithfully represents} $\tau$ under $\Delta$ if, for all $v,w\in\mathcal V_\tau$,
\[
\begin{aligned}
  \W^{\Delta}_{\ell}(v)&\ne\varnothing
    &&\text{(totality)},\\
  v\ne w&\implies\W^{\Delta}_{\ell}(v)\cap\W^{\Delta}_{\ell}(w)=\varnothing
    &&\text{(absolute injectivity)}.
\end{aligned}
\]
\end{definition}

Notably, absolute injectivity is analogous to the ``no confusion'' property of~\citet{diatchki2005high}. Since $\W^{\Delta}_{\ell}$ is a relation, one value may have several representations, and absolute injectivity only forbids two values from sharing one. A representation $\eta$ \emph{determines} $v$ when it represents $v$ and no other value of its type. Under a faithful layout, each permitted representation determines its source value. We write $\upsdW^{\Delta}_{\ell}(\eta)=v$ for this partial recovery function, reminiscent of the representation function of~\citet{hoare1989proof}.

\subsection{Well-Formedness}
\label{sec:well-formedness}

Injectivity alone cannot keep values from being lost, e.g., pruning the $\Ctor{Leaf}$ arm from the body of $\RefT{Tree}$ maintains absolute injectivity, yet every tree ends in leaves, so every tree loses its representation. The judgment $A\vdash\ell\;\mathsf{wf}$ establishes totaliy syntactically, checking each component only on the subjects that reach it. Its rules, defined in \Aref{app:well-formedness}, require, e.g.,  that field selections are defined, memoized expressions fit their widths, and sum arms cover every admitted subject.

A reference, on the other hand, defers its check to the corresponding layout definition. The environment $\Delta$ is \emph{well formed} when each body is well formed on all values of its type, $\mathcal V_{\RefT T}\vdash\Delta(\RefT T)\;\mathsf{wf}$. We require layout definitions to be \emph{structurally recursive}, so each reference applies to a smaller source value. Syntactically, every reference in a body must occur within a named layout whose path begins with a field name, e.g., $\mathsf l:\RefL{Tree}$ applies the $\RefL{Tree}$ layout to the left child. We package the assumptions as one invariant, $\mathsf{Inv}$, and prove that well-formedness suffices for totality.

\begin{definition}[Environment invariant \lean{Core.Inv}]
\label[definition]{def:rewrite-invariant} We write $\mathsf{Inv}(\Delta)$ when $\Delta$ is well formed and structurally recursive.
\end{definition}

\begin{lemma}[Well-formedness implies totality \lean{WF.totality}]
\label[lemma]{lem:totality}
If $\mathsf{Inv}(\Delta)$ and $A\vdash\ell\;\mathsf{wf}$, then $\W^{\Delta}_{\ell}(v)\ne\varnothing$ for every $v\in A$.
\end{lemma}

\section{Correctness of Layout Rewrites}
\label{sec:rewrite-correctness}

We prove each rewrite locally, then lift its guarantee through the surrounding layout and recursive definitions. The local requirement compares the focus before and after the rewrite. Importantly, a focus need not faithfully represent its whole subject, e.g., the field $\mathsf l:\RefL{Tree}$ alone cannot distinguish nodes with equal left children but different keys.

\subsection{Local Refinement}
\label{sec:local-reconstruction}

Write $v\approx^{\Delta}_{\ell}w$ when $v$ and $w$ have a common representation under $\ell$, that is, $\W^{\Delta}_{\ell}(v)\cap\W^{\Delta}_{\ell}(w)\ne\varnothing$. Layout $\ell'$ under $\Delta'$ \emph{refines} $\ell$ under $\Delta$ on subjects $A$ when
\[
  \forall v,w\in A.\quad
  v\approx^{\Delta'}_{\ell'}w
  \implies v\approx^{\Delta}_{\ell}w.
\]
Refinement thus preserves absolute injectivity.

\begin{definition}[Local refinement \lean{Core.LocalRefinement}]
\label[definition]{def:local}
Under $\Delta$, layout $\ell_1$ \emph{locally refines} $\ell_0$ on subjects $A$, written $\ell_0\preceq\ell_1$, if both layouts are well formed on $A$ and $\ell_1$ refines $\ell_0$ on $A$.
\end{definition}

Each primitive rule proves refinement, a reflexive and transitive relation, with a \emph{reconstruction function} $\mathsf f:\IR\rightharpoonup\IR$. For every $v\in A$ and $\eta'\in\W^{\Delta}_{\ell_1}(v)$, it returns $\mathsf f(\eta')\in\W^{\Delta}_{\ell_0}(v)$. Refinement then follows at once, i.e., if $v$ and $w$ share $\eta'$ after the rewrite, they share $\mathsf{f}(\eta')$ before it. For example, padding reconstructs by dropping the added component, commutation by swapping the pair, and $\mathsf{split}_k$ by joining its two scalars as $(b_0,b_1)\mapsto b_0+2^kb_1$. Composing a derivation's reconstruction functions with the canonical decoder gives a decoder for the rewritten layout, although such a decoder may be inefficient if naively implemented.

\subsection{Lifting Local Refinement}
\label{sec:context-replacement}
\label{sec:faithfulness-under-rewriting}

Since a rewrite leaves the rest of the layout unchanged, the distinctions that local refinement preserves at the focus carry over to the whole layout. Updating a layout definition is subtler: local refinement is proved with the old body at recursive references, but the update changes those references too.

\begin{theorem}[Contextual replacement \lean{Core.LocalRefinement.context}]
\label[theorem]{thm:contextual-replacement}
Let $\Ctx[\ell_0]$ be well formed on $B$, and let $A$ be the subjects that $\Ctx$ assigns to its hole from $B$. If $\ell_0\preceq\ell_1$ on $A$, then $\Ctx[\ell_0]\preceq \Ctx[\ell_1]$ on $B$.
\end{theorem}

\begin{lemma}[Recursive binding \lean{Core.LocalRefinement.bind}]
\label[lemma]{lem:recursive}
Let $\Delta\;\mathsf{wf}$, $\Delta(\RefT T)=\ell_0$, and $\Delta'=\Delta[\RefT T\mapsto\ell_1]$.
If $\ell_0\preceq\ell_1$ on $\mathcal V_{\RefT T}$ under $\Delta$ and $\ell$ is well formed on $B$, then $\ell$ under $\Delta'$ refines $\ell$ under $\Delta$ on $B$.
\end{lemma}

The proof inducts on one subject's representation under $\Delta'$ and compares the other subject's derivation of the same representation. At a reference to $\RefT T$, the induction hypothesis yields a shared representation of $\ell_1$ under $\Delta$, and local refinement gives a shared representation of $\ell_0$. A \emph{valid rewrite} satisfies local refinement at its focus and either replaces that focus under the premises of \Cref{thm:contextual-replacement}, or updates a definition under the premises of \Cref{lem:recursive} while preserving structural recursion.

\begin{theorem}[Soundness of local rewrites \lean{Core.Rewrite.sound}]
\label[theorem]{thm:rewrite-soundness}
Suppose $\mathsf{Inv}(\Delta)$ and $A\vdash\ell\;\mathsf{wf}$ under $\Delta$. Every valid rewrite from $(\Delta,\ell)$ to $(\Delta',\ell')$ establishes $\mathsf{Inv}(\Delta')$ and $A\vdash\ell'\;\mathsf{wf}$ under $\Delta'$, and $\ell'$ under $\Delta'$ refines $\ell$ under $\Delta$ on $A$.
\end{theorem}

If $\ell$ faithfully represents $\tau$ and $A=\mathcal V_\tau$, the preserved invariant and well-formedness give totality by \Cref{lem:totality}, and refinement preserves absolute injectivity. Thus the new layout remains faithful. By transitivity, every layout reached from a canonical layout by valid rewrites is faithful.

\subsection{Adding a Rewrite Rule}
\label{sec:extending-rewrites}

The rewrite rules of \Cref{fig:rewrite-rules} are sound but not complete, i.e., not every \emph{faithful} layout can be derived. The rule set can be extended to express novel layouts, provided each new rule establishes local refinement (\Cref{def:local}) with a reconstruction function. \Cref{thm:rewrite-soundness,thm:all-rule-program-preservation} then apply to it unchanged. As a concrete example, we realize an optimization used in Scion for storing an interval as an offset and a count~\cite[Fig.~5]{gyurgyik2026decoupling}. When $k<n$, an interval $[\mathsf{lo},\mathsf{hi})$ can store its start and length,
\[
  \mathsf{lo}:\Uint n\times\mathsf{hi}:\Uint n
  \longrightarrow
  \mathsf{lo}:\Uint n\times\Stored{(\mathsf{hi}-\mathsf{lo}):\Uint k} \;\;\; \text{\lean{Core.Interval.rewrite}}
\]
on admitted subjects with $\mathsf{lo}\le\mathsf{hi}<2^n$ and $\mathsf{hi}-\mathsf{lo}<2^k$. The reconstruction $\mathsf f(b,d)=(b,b+d)$ recovers the original pair, and, together with well-formedness, this establishes local refinement.

\subsection{Program Preservation}
\label{sec:program-preservation}

Layout rewrites should change how programs represent data while preserving what they compute. We prove this for a pure language with constructors, matching, let bindings, and recursive calls, under call-by-value big-step semantics~\cite{kahn1987natural}. We compare terminating programs by their scalar results, the observations on which all layouts agree (details in~\Aref{app:program-preservation}); this does not verify generated code.

A program evaluates over source values, $P;\rho\vdash e\Downarrow v$, or over represented values, $\Delta,b,\iota;P;\rho'\vdash e\Downarrow r$, where a construction policy $b$ picks a representation for each new value and a destruction policy $\iota$ reads one back. The policies are \emph{valid} when $b$ succeeds exactly on values of the requested type and returns a permitted representation, and $\iota$ recovers each such value from every permitted representation. We write $r\sim_\Delta v$ when $r$ represents $v$, relate environments that bind the same variables to related values, and call $\Delta$ \emph{faithful} when $\RefL T$ faithfully represents $\RefT T$ for every type $\RefT T$.

\begin{theorem}[Representation independence \lean{Programs.representation_independence}]
\label[theorem]{thm:representation-independence}
Suppose $\Delta_0$ and $\Delta_1$ are faithful. For valid policies $(b_0,\iota_0)$ at $\Delta_0$ and $(b_1,\iota_1)$ at $\Delta_1$ and related inputs
$\rho_0\sim_{\Delta_0}\rho$ and $\rho_1\sim_{\Delta_1}\rho$, every program $P$, expression $e$, and scalar $k\in\mathbb N$ satisfies
\begin{equation}
\label{eq:program-observations}
    \Delta_0,b_0,\iota_0;P;\rho_0\vdash e\Downarrow k
    \quad\Longleftrightarrow\quad
    \Delta_1,b_1,\iota_1;P;\rho_1\vdash e\Downarrow k.
\end{equation}
\end{theorem}

\begin{theorem}[Program preservation \lean{Core.BodyRewriteSeq.{observations,roots}}]
\label[theorem]{thm:all-rule-program-preservation}
Suppose $\mathsf{Inv}(\Delta_0)$, $\Delta_0$ binds every declared type, and $\Delta_0$ is faithful. If $\Delta_0\longrightarrow^*\Delta_1$ is a finite sequence of valid updates to layout definitions, then $\Delta_1$ is faithful, and \eqref{eq:program-observations} holds for all valid policies at the two endpoints and all related inputs.
\end{theorem}

For rewritten layouts, the final layout yields a valid construction policy $b$ (under \Cref{lem:totality}). A valid destruction policy $\iota$ is recovered by inverting the rewrite rules through their reconstruction functions (\Cref{sec:local-reconstruction}).

\section{From Structural Representation to Bits}
\label{sec:physical-storage}
Structural representations retain component boundaries and pointee contents, but a bit-level encoding that discards this information can be ambiguous even when the layout is faithful. Recovery from memory therefore requires a separate guarantee from an encoder and decoder, together called a \emph{codec}. Prior work has studied the construction and verification of binary codecs extensively~\cite{geest2017generic,delaware2019narcissus,bangert2014nail,ye2019protocol,fisher2005pads}. We connect this work to layout rewriting by stating the recovery property a codec must satisfy, then give a simple codec satisfying it under explicit premises.

\subsection{Recovery from Physical Storage}
\label{sec:storage-contract}
\label{sec:forward-encoder-decoder}

A physical encoding $m=(\beta,H)\in\mathcal M$ consists of a finite top-level bit vector $\beta$ and a finite heap $H$ from addresses to bit vectors. For a layout $\ell$ under $\Delta$, an encoder and a decoder are partial functions
\[
  \mathsf{encode}_{\ell}:\IR\rightharpoonup\mathcal M,
  \qquad
  \mathsf{decode}_{\ell}:\mathcal M\rightharpoonup\IR,
\]
and we leave $\Delta$ implicit in their notation. The \emph{storage recovery property} requires decoding to preserve the represented source value. For all $v\in\mathcal V_\tau$, $\eta\in\IR$, and $m\in\mathcal M$,
\begin{equation}
\label{eq:storage-round-trip}
\begin{aligned}
  \eta\in\W^{\Delta}_{\ell}(v)
  \ \land\ \mathsf{encode}_{\ell}(\eta)=m
  \quad\Longrightarrow\quad{}
  \exists\eta'\in\W^{\Delta}_{\ell}(v).\quad
  \mathsf{decode}_{\ell}(m)=\eta'.&
\end{aligned}
\end{equation}
The decoder receives only the memory; it may return another representation of $v$, so recovery does not require identical padding, and it is unconstrained on memories that no encoding produces.

\begin{theorem}[Recovery after rewriting and encoding \lean{Core.storage_recovery}]
\label[theorem]{thm:storage-recovery}
Suppose $\mathsf{Inv}(\Delta)$ and $\ell$ is well formed on $\mathcal V_\tau$ and faithfully represents $\tau$ under $\Delta$. Let $(\Delta',\ell')$ be obtained by a finite sequence of valid rewrites. If an encoder and a decoder for $\ell'$ under $\Delta'$ satisfy \Cref{eq:storage-round-trip}, then for every $v\in\mathcal V_\tau$, $\eta\in\W^{\Delta'}_{\ell'}(v)$, and successful encoding $\mathsf{encode}_{\ell'}(\eta)=m$, the decoder returns a representation that determines $v$, so $\upsdW^{\Delta'}_{\ell'}(\mathsf{decode}_{\ell'}(m))=v$.
\end{theorem}

The theorem guarantees recovery of the source value provided that encoding succeeded. For an end-to-end instance, the mechanization supplies a simple forward encoder and decoder, which place scalars, products, and pointers in a bit vector and heap and reads them back left to right. Encoding succeeds when every ordinary sum has been tagged, every tag field is stored whole in its scope, and there are enough addresses for the pointers. Decoding inverts every successful encoding when, in addition, each tag field precedes its payload. Recovery of the source value then follows from \Cref{thm:storage-recovery}; the mechanization carries this out for Layout~\ref{lst:low-bits}, and \Aref{app:physical-codec} elaborates on the conditions.

\subsection{Decoding Limitations}
\label{sec:decoding-limitations}

Recovery from storage also depends on factors beyond faithfulness. For example, commuting a tag after the arm it selects preserves faithfulness, but the forward decoder is required to read the arm before learning which arm it is. On the other hand, a backward decoder, i.e., one that reads fields in reverse order, could still recover the values. Moreover, variable-width arms can make the encoding itself ambiguous. Consider the type $\RefT S=\Ctor A(x:\Uint1)\mid\Ctor B(z:\Uint2)$ with initial layout $t:\Uint1\times\Tagged t{\ar{(0,\Ctor A)}{x:\Uint1},\ \ar{(1,\Ctor B)}{z:\Uint2}}$, and then store a pair of $\RefT S$ values with the first component's tag after its payload and the second component's tag before. Then,
\[
\begin{aligned}
  (\Ctor A(1),\Ctor B(1))&\longmapsto\mathtt{10}\mathbin{\Vert}\mathtt{101}=\mathtt{10101},\\
  (\Ctor B(2),\Ctor A(1))&\longmapsto\mathtt{101}\mathbin{\Vert}\mathtt{01}=\mathtt{10101}
\end{aligned}
\]
so no decoder can recover both pairs. Both tags lie at the boundary between the components, but only a tag reveals where that boundary is. Narcissus resolves such ambiguities by fixing boundaries from the surrounding format~\cite[Sec.~3.1]{delaware2019narcissus}. Local rewrites thus preserve faithfulness but not physical recoverability, which depends on the layout as a whole.

\section{The \sys{} Programming Language}
\label{sec:implementation}

The preceding sections define how a layout organizes data in memory. The compiler must abide by this specification, but, importantly, is free to choose how construction, destruction, and memory management are realized. We next describe how \sys{} exploits this to generate efficient code.

\subsection{Core Language and Layout Selection}
\label{sec:core-language}

\begin{figure}[htbp]
\centering
\small
\(
\begin{array}{@{}r@{\;}c@{\;}l@{\qquad}r@{\;}c@{\;}l@{\qquad}r@{\;}c@{\;}l@{}}
\multicolumn{9}{c}{x,f,\RefT T,\Ctor K\in\mathsf{Id}\qquad c\in\mathsf{Lit}\qquad n\in\mathbb{N}\qquad t\in\{\mathsf{u}_n,\mathsf{i}_n,\mathsf{f}_n\}}\\[0.6ex]
P &::=& \overline d\;\overline F
 &F&::=& \mathtt{fn}\ f(\overline{x:\tau}):\tau\mathrel{\mathtt{=}}e
 &e&::=& x\mid c\mid\Ctor K(\overline e)\mid f(\overline e)\mid e_1 \oplus e_2\\
d &::=& \mathtt{data}\ \RefT T
        \mathrel{\mathtt{=}} \kappa_1\mathrel{\mathtt{|}}\cdots
        \mathrel{\mathtt{|}}\kappa_n\mathtt{;}
 &\tau&::=& t\mid\RefT T\mid[x]\tau
 &&\mid&\mathtt{let}\ x:\tau\mathrel{\mathtt{=}}e_1\;\mathtt{in}\;e_2\\
\kappa &::=& \Ctor K(\overline{x:\tau})
 &p&::=& x\mid\mathtt{\_}\mid n\mid\Ctor K(\overline p)
 &&\mid&\mathtt{match}\ e\;\{\overline{p\Rightarrow e}\}
\end{array}
\)
\caption{Monomorphic, first-order core of \sys{}.}
\label{fig:source-grammar}
\end{figure}

\Cref{fig:source-grammar} gives the grammar of the monomorphic, first-order core of \sys{}. The frontend monomorphizes generic definitions and defunctionalizes closures into ADTs, so every value has a concrete type. Scalars are signed integers, unsigned integers, and floating-point numbers. Variable-length arrays $[x]\tau$ are first-class, with an in-scope scalar length $x$~\cite{iso1999c,paszke2021getting}. The formalism models a finite array as a sum with one constructor per length, indexed by its length; the implementation provides an explicit array construct for simplicity.

For types without rewrites applied, \sys{} uses a \emph{default} layout. Fields keep their declared order, fixed-size fields are inlined, recursively defined fields sit behind a pointer, and tags are rounded up to whole bytes. 
When a sum is stored inline within another value its arms are padded to a common size, so that following fields have statically known offsets. For heap-allocated values, we size cells per constructor rather than padding them to a common size; each cell includes a 32-bit reference count. \Cref{sec:evaluation-comparison,sec:dse} show that this simple policy strikes a balance between layout compactness and runtime performance.

\subsection{Layout-Directed Compilation}
\label{sec:layout-compilation}

\paragraph{Well-formedness} Before generating code, the compiler verifies the machine can realize each layout. This \emph{machine well-formedness} requires that tags and array lengths are recoverable, fields admit an acyclic reading order satisfying their offset dependencies, every cycle in the storage graph crosses a pointer, and storage sizes meet allocation constraints. Pointer compression, for example, must comply with the requirements of the target architecture. These conditions are in part intended to exclude the ambiguous encodings of \Cref{sec:decoding-limitations}.

\paragraph{Construction and destruction} Unlike layout description languages~\cite{baudon2023bitstealing, gyurgyik2026decoupling, chen2023dargent} which decode a value by inverting its layout specification, \sys{} generates code by following a layout's rewrites from a canonical layout. Each rewrite updates two artifacts, (1) the layout's \emph{decoder}, a function from a value's stored form to the source value, and (2) a \emph{storage plan} that records each stored component's physical metadata, e.g., its offset and width. Construction proceeds by reading the storage plan, while destruction, i.e., pattern matching and field access, queries the decoder. A new rule thus requires supplying only its reconstruction function and its changes to the storage plan.

\paragraph{Producer and consumer specialization} Compiling pattern matches and constructions field by field, as the source writes them, can make a program asymptotically slower under a rewritten layout. Each field the source binds or supplies is a value of its own, while the layout may keep that value only inside a larger component. For example, when an unfolded list stores two elements per cell, matching \code{Cons(h, t)} binds \code{t} to a tail with no standalone representation. Rebuilding that tail means repacking the entire suffix, and doing so at every step turns a linear traversal quadratic. \sys{} instead specializes consumers and producers to the layout through a source-to-source translation that combines deforestation~\cite{wadler1990deforestation} and partial evaluation~\cite{jones1993partial} techniques to avoid this asymptotic slowdown when possible. The algorithms are discussed further in \Aref{app:specialization}.

Specialized pattern matching is then lowered to decision trees~\cite{maranget2008compiling} whose tests read physical encodings, and branch provenance removes tests whose outcome is already known. Unlike the formalism, the implementation treats the arms of a sum as an ordered list, and a rule that commutes arms lets programmers choose which guard construction tests first\footnote{We further explored decoupled programmer control of decision-tree compilation but saw little runtime benefit for our workloads, consistent with~\citet{scott2000match}. \sys{} defaults to the necessity heuristic of~\citet{maranget2008compiling}.}.

\paragraph{Instruction selection} Naively executing a decoder for every query would rebuild an entire source value just to read a single field. The compiler instead optimizes the decoder, using the storage plan to simplify instruction selection. Knowledge of field positions and widths lets the compiler combine representation operations with the instructions that use their results. For example, once a branch establishes that a pointer carries the low-bit tag $t$, untagging folds into each field access, as in GHC's pointer tagging~\cite{marlow2007faster}, so a field at byte offset $d$ from the untagged pointer is read at offset $d-t$ from the tagged word, without a separate mask instruction. Finally, \sys{} lowers the specialized program to SSA~\cite{rosen1988global,cytron1991efficiently}, applies standard compiler optimizations~\cite{pettersson1992term,aho2006compilers,leijen2023trmc,cocke1970global}, and emits LLVM IR~\cite{lattner2004llvm}.

\subsection{Reference Counting}
\label{sec:reference-count}

\sys{} manages memory with ownership-based reference counting, following the compiler-based approach of prior work~\cite{ullrich2019counting, reinking2021perceus, lorenzen2023fp2}. In \sys{}, however, the count is itself part of the layout. The rewrite rule \code{rc n} prepends an unsigned $n$-bit count to the focused layout, a field that the compiler manages but other rewrites treat like any other, so a count can be narrowed, moved, or removed entirely. Omitting the count asserts that each instance of the type has a single owner, which \sys{} verifies after reference count optimizations are applied. As in Koka~\cite[Sec.~2.7.2]{reinking2021perceus}, counts are \emph{sticky}, so an $n$-bit count saturates once its high bit is set, and its cell is then neither reclaimed nor reused.

Layout control lets programmers choose cell sizes that enable reuse and omit reference counts when ownership permits. Dynamic reuse (\code{rc}-1) updates a dying cell in place when its count is one at run time~\cite{reinking2021perceus, brandon2026fullyautomatic}. It enables cells of the same size to be reused between constructors of the same or different types, and arrays with a single owner to be updated in place. Layout rewrites let programmers choose between per-arm allocation sizes and a common cell size. Per-arm sizes avoid padding smaller arms, while padding all arms to a common size enables cell reuse across all constructors of the type. Additionally, the programmer can entirely omit the reference count field from a layout, which the compiler accepts only when static ownership analysis verifies unique ownership; we call this optimization \code{rc}-0. The $k$-d tree case study, the layout design space exploration, and the Ribbit comparison exercise \code{rc}-0 (\Cref{sec:evaluation-case-studies,sec:dse,sec:evaluation-ribbit}).

\subsection{Layout Coercion}
One layout rarely suits every phase of a program, since construction and traversal can favor different representations of the same ADT. The scheduling directive \code{coerce} changes a value's representation at a chosen binding without changing the program. Logically, the conversion is a catamorphism $h$~\cite{meijer1991functional} that converts a value's children and rebuilds each constructor in the target representation,
\[
\begin{aligned}
  h\bigl(\Ctor K_0(\bar a,t_1,\ldots,t_k)\bigr) &=\Ctor K_1\bigl(\bar a,h(t_1),\ldots,h(t_k)\bigr).
\end{aligned}
\]
Here $\bar a$ are the nonrecursive fields, and subscripts $0$ and $1$ mark the source and target representations. Because both sides share one ADT, the fold's constructor functions are fixed, and the compiler implements them by specializing construction to the target layout and matching to the source layout. Conversion thus reuses the machinery of \Cref{sec:layout-compilation}. The 2-3 tree and $k$-d tree case studies evaluate \code{coerce} between their construction and query phases (\Cref{sec:evaluation-case-studies}).

\subsection{Using Memoized Expressions}
A layout can also store what a program would otherwise recompute. A memoized expression moves work from queries to construction. For example, given the type $\RefT{T}=\Ctor{T}(a:\mathsf i_{32},b:\mathsf i_{32})$, the rewrite \code{memoize (a + b)} stores the sum of fields \code{a} and \code{b} in a new field, which the generated constructors compute. During destruction, the compiler annotates each cache load with the expression it represents. After canonicalization, common-subexpression elimination (CSE)~\cite{cocke1970global} replaces matching computations with the cached value. Memoized expressions may also be recursive; the interval-tree case study (\Cref{sec:evaluation-case-studies}) evaluates such an instance.

\setcounter{topnumber}{1}
\setcounter{bottomnumber}{1}
\setcounter{totalnumber}{2}
\section{Evaluation}
\label{sec:evaluation}

Our evaluation demonstrates that \sys{} achieves competitive performance against existing implementations using automatically selected default layouts, and that composing rewrites can improve runtime and reduce peak memory further. Layout control also complements automatic memory management by shrinking allocations that reuse does not eliminate. We substantiate these findings with measurements of runtime and peak memory.

\subsection{Experimental Methodology}
\label{sec:methodology}


All benchmarks run single-threaded on an Apple M2 Pro with \ResultCPUCores{} cores and \ResultMemoryGiB{}\,GiB of memory, running macOS~\ResultMacOS{}. \sys{} emits LLVM IR, which Clang~\ResultClang{} compiles with \code{-O3 -mcpu=native}. We compare against MLton~\cite{weeks2006whole}, Morphic~\cite{brandon2023better,brandon2026fullyautomatic}, Ribbit~\cite{baudon2024thesis,baudon2023bitstealing}, and Koka~\cite{leijen2014koka,lorenzen2023fp2}\footnote{MLton \texttt{on-20241230-release} (\url{https://github.com/MLton/mlton}), Morphic \texttt{da679c4} (\url{https://github.com/morphic-lang/morphic}), Ribbit \texttt{35a2225} (\url{https://gitlab.inria.fr/ribbit/ribbit}), and Koka \texttt{v\ResultKoka{}} (\url{https://github.com/koka-lang/koka})}. Every timed system measures wall-clock time, and each configuration runs in fresh processes, with \ResultWarmups{} discarded warm-ups followed by \ResultSamples{} measured runs. We report the mean of the \ResultRetainedSamples{} runs left after discarding the \ResultTrimEachTail{} fastest and \ResultTrimEachTail{} slowest, with approximate \ResultConfidencePercent{}\% error bars. For memory, we report the median peak resident set size (RSS). Unless noted, we use mimalloc~\ResultMimalloc{}~\cite{leijen2019mimalloc}, and~\Aref{app:system-results} repeats the results with the system allocator. We verify every configuration computes the same answer and \sys{} has no memory leaks. Experiments below describe any remaining setup.

\noindent \textit{AI Use}. AI~\cite{anthropic_software, openai_software} assisted with implementation of the \sys{} compiler from an initial prototype, proof mechanization, the benchmark harness, paper figures, and the layout search of \Cref{sec:dse}.

\subsection{Case Studies}
\label{sec:evaluation-case-studies}
Three case studies measure what layout rewrites gain in runtime and memory within a single program, by composing successive rewrites on a \textit{k}-d tree, memoizing an expression in an interval tree, and changing layouts between the insertions and range scans of a 2-3 tree.

\casestudyheading{Composing Rewrites on \textit{k}-d Trees}{\KdTreeType}
\begin{figure}[!t]
  \centering
  \captionsetup{skip=0pt}
  \includegraphics[width=\linewidth]{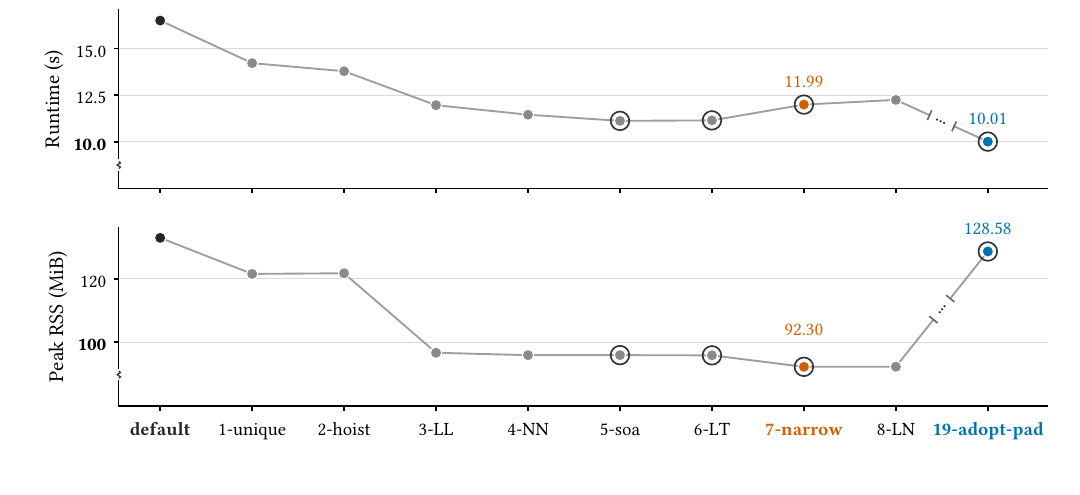}
  \caption{Lower is better. Runtime and peak RSS across $k$-d tree layouts. The line denotes successive composition of rewrite rules. \KDMinRuntimeMarker{} marks minimum runtime, \KDMinRSSMarker{} minimum RSS, and \KDParetoMarker{} the observed Pareto frontier.}
  \label{fig:evaluation-kd-tree}
\end{figure}

We explore the interaction of layout rewrites through two-point correlation counting~\cite{moore2003fast}. The program counts points within a query radius, using a $k$-d tree~\cite{bentley1975multidimensional} to accelerate the query. We build a tree over \ResultKDPoints{} points and issue \ResultKDQueries{} queries. Starting from \sys{}'s default layout, each configuration in \Cref{fig:evaluation-kd-tree} adds one optimization to the layout before it, realized through a short sequence of rewrites. The results show that effective layouts can be reached incrementally by composing simple rewrites, and additionally expose a tradeoff between runtime performance and memory use.

Because the tree is built once and then only read, \code{1-unique} can drop the reference count, which ownership analysis confirms no cell needs (\code{rc}-0, \Cref{sec:reference-count}), and every leaf moves to an allocation bin half the size. \code{2-hoist} then moves the tag into low bits of each pointer~\cite{baudon2023bitstealing, elsman2024double,weeks2006whole,marlow2007faster}, so a traversal dispatches on bits it already holds instead of loading a byte, and a leaf is just the three coordinates. Next, most nodes near the bottom of the tree have two leaf children, and \code{3-LL} unfolds both into their parent~\cite{benthin2018compressed,shao1994unrolling, hall1994using}, turning three cells into one 48-byte cell and eliminating two pointer dereferences. \code{4-NN} unfolds two $\Ctor{Node}$ children the same way, which barely moves either metric but places two levels of the tree in one cell for later steps to reshape. \code{5-soa} then reorders the inlined leaf coordinates and the bounding-box fields for vectorized arithmetic~\cite{woop2014exploiting, benthin2018compressed, ylitie2017efficient}, so one 2-wide comparison tests both leaves, and \code{6-LT} inlines a lone $\Ctor{Leaf}$ child while keeping its sibling behind a pointer. The last two incremental steps trade runtime for memory. \code{7-narrow} refines pointers to 48 bits, moving two-$\Ctor{Node}$ cells from the 112- to the 96-byte allocation bin~\cite{chilimbi1999cache, lattner2005transparent} at the cost of a slower load for every child, and \code{8-LN} also inlines the $\Ctor{Node}$ sibling of a lone leaf.

The final layout \code{19-adopt-pad} performs a \code{coerce} step (included in the timed region) between the construction and query phases. This scheduling directive builds the tree under a layout that keeps children boxed and steals bits for the tag, then converts it to an unfolded, alignment-friendly layout for better instruction selection during the queries. It also builds the point buffer's array inline in the counted cell, and then pads leaves in the read layout within their existing allocation bin. Of the layouts shown, \texttt{\ResultKDFastestLayout{}} is the fastest, $\ResultKDSpeedup{}\times$ faster than the default, and \texttt{\ResultKDSmallestLayout{}} has the smallest peak RSS, \ResultKDRSSReduction{}\% below the default's. Importantly, the fastest and smallest layouts differ, thus no single layout is optimal across both runtime and peak memory use.

\casestudyheading{Memoized Expressions in Interval Trees}{\IntervalTreeType}
An interval-tree query uses the maximum upper endpoint in a subtree to choose its next branch~\cite[Sec.~15.3]{cormen1990introduction}. An unaugmented tree recomputes this maximum by traversing the left subtree of each node that a query visits, so a query can take time linear in the number of intervals. Augmenting the maximum at every tree node instead makes a query take time proportional to the tree's height. The cached maximum is defined recursively,
\[
\mu(t) =
\begin{cases}
0, & t \in \Ctor{Leaf} \\
\max\bigl(\mu(\mathsf{l}), \mu(\mathsf{r}), \mathsf{hi}\bigr), & t \in \Ctor{Node}(\mathsf{l}, \mathsf{r}, \mathsf{lo}, \mathsf{hi}).
\end{cases}
\]
Memoized expressions cache this value directly in the ADT, analogous to the reduction \emph{annotations} in Bonsai~\cite[Sec. 4]{root2026bonsai}. \sys{} realizes this augmentation through layout rewrites over the unaugmented source. MLton and Morphic reach the same asymptotic cost only by changing the source, so all three systems also run a hand-augmented program that stores the maximum in every constructor. We time the queries alone, issuing \ResultIntervalQueries{} point queries to trees of $2^n$ intervals for $n \in [\ResultIntervalExponentStart{},\ResultIntervalExponentStop{})$, with the compiler configurations of \Cref{sec:evaluation-comparison}. Across all tree sizes, the memoized \sys{} layout is \ResultIntervalMorphicSpeedup{}$\times$ faster than hand-augmented Morphic and \ResultIntervalMLtonSpeedup{}$\times$ faster than hand-augmented MLton (geomean) as shown in \Cref{fig:evaluation-interval-tree} (left). The memoized layout also performs similarly to the hand-augmented \sys{} program at every size, showing that expressing the augmentation as a layout rewrite adds no overhead.

\begin{figure}[!t]
  \centering
  \captionsetup{skip=2pt}
  \begin{subfigure}[t]{0.48\linewidth}
    \vspace{0pt}
    \centering
    \includegraphics[width=\linewidth,trim=0 7 0 0,clip]{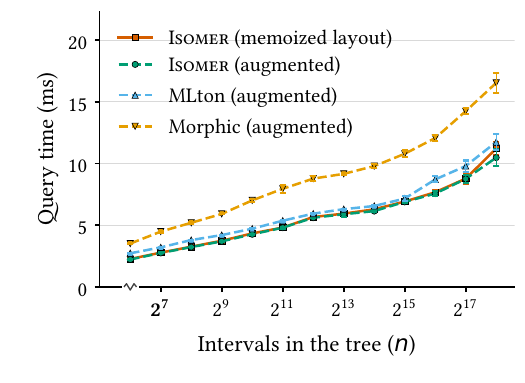}
  \end{subfigure}\hfill
  \begin{subfigure}[t]{0.48\linewidth}
    \vspace{0pt}
    \centering
    \includegraphics[width=\linewidth,trim=0 7 0 0,clip]{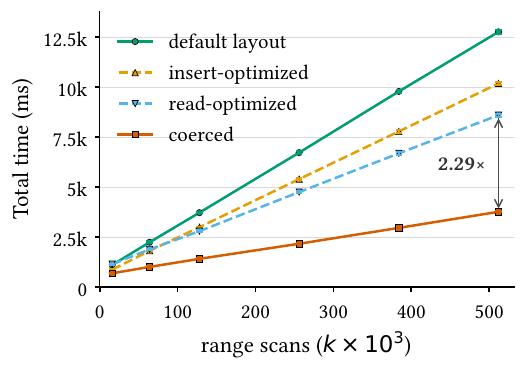}
  \end{subfigure}
  \caption{Lower is better. Interval-tree query time with memoized expressions in \sys{} and with a hand-augmented source in \sys{}, MLton, and Morphic (left), and total runtime for $2$--$3$ tree insertion and range scans, including data structure translation (right).}
  \label{fig:evaluation-tree-workloads}
  \label{fig:evaluation-interval-tree}
  \label{fig:evaluation-tree23}
\end{figure}

\casestudyheading{Changing Layouts in 2-3 Trees}{\TwoThreeTreeType}
The interval-tree cache moves recursive work from queries to construction; a complementary opportunity arises when construction and traversal favor different representations. The \code{coerce} directive lets a program convert once between phases and use a compact layout for repeated scans. A 2-3 tree~\cite[Sec.~6.2.4]{knuth1998sorting} is a balanced search tree whose internal nodes hold one or two keys and have two or three children, respectively, with all leaves at the same depth. We insert \ResultTreeKeys{} keys, then issue \ResultTreeScanCounts{} range scans for $k\in\ResultTreeScanMultipliers$ and a final reduction. \Cref{fig:evaluation-tree23}~(right) compares the default layout with three configurations. \code{insert-optimized} pads $\Ctor{Two}$ cells to match $\Ctor{Three}$ cells for reuse, \code{read-optimized} inlines empty $\Ctor{Leaf}$ children in leaf-level nodes, and \code{coerced} builds with \code{insert-optimized} and then translates via \code{coerce} to \code{read-optimized} before scanning. The best fixed layout depends on the workload, switching from \code{insert-optimized} to \code{read-optimized} between 64{,}000 and 128{,}000 scans. \code{coerced} pays for one conversion but is faster than both throughout the sweep. At \ResultTreeMaxScans{} scans, it is $\ResultTreeBestFixedSpeedup{}\times$ faster than the best fixed layout. The \code{coerce} directive thus turns layout choice from a program-wide decision~\cite{gyurgyik2026decoupling} into a phase-specific one.

\subsection{Comparison with Monomorphizing Languages}
\label{sec:evaluation-comparison}

MLton and Morphic are whole-program compilers that monomorphize and, unlike \sys{}, do not offer user control of layouts. Morphic computes strongly connected components (SCCs) over datatype-size dependencies and boxes an arm when its payload refers to a type in its own SCC. MLton's representation emerges from interacting whole-program passes, including monomorphization, unboxing, argument flattening, and representation selection~\cite{weeks2006whole}. \Cref{fig:evaluation-tmap-layouts} shows the layout each system chooses for \code{tmap}'s \code{Tree} ADT. Morphic opts for aligned fields at the cost of unused bits, while MLton employs more advanced layouts automatically, e.g., bit stealing techniques.

\begin{figure}[!b]
\centering
\footnotesize
$\RefT{Tree}=\Ctor{Leaf}\mid\Ctor{Bin}(l:\RefT{Tree},x:\mathsf i_{32},r:\RefT{Tree})$\par
\vspace{4pt}
\begingroup
\scriptsize
\thinmuskip=1mu
\medmuskip=1mu
\thickmuskip=2mu
\begin{tabular}{@{}l@{\quad}l@{}}
{\small Morphic} &
$t:\Uint1\times\Zero7\times\Pad{56}\times
  \Tagged t{\ar{(0,\Ctor{Leaf})}{\Pad{64}},\ \ar{(1,\Ctor{Bin})}{\Ptr{64}{\mathsf{rc}:\Uint{64}\times l:\RefL{Tree}\times x:\mathsf i_{32}\times\Pad{32}\times r:\RefL{Tree}}}}$
\\[4pt]
{\small MLton} &
$t:\Uint2\times
  \Tagged t{\ar{(1,\Ctor{Leaf})}{\Zero{62}},\ \ar{(0,\Ctor{Bin})}{\Ptr{62}{\mathsf{gc\_metadata}:\Uint{64}\times x:\mathsf i_{64}\times l:\RefL{Tree}\times r:\RefL{Tree}}}}$
\\[4pt]
\small{\sys{}} (default) &
$\Ptr{64}{\mathsf{rc}:\Uint{32}\times t:\Uint8\times
  \Tagged t{\ar{(0,\Ctor{Leaf})}\epsilon,\ \ar{(1,\Ctor{Bin})}{l:\RefL{Tree}\times x:\mathsf i_{32}\times r:\RefL{Tree}}}}$
\\[4pt]
\small{\sys{}} (optimized) &
$\Tagged t{\ar{(0,\Ctor{Leaf})}{\Pad{56}},\ \ar{(1,\Ctor{Bin})}{\Ptr{56}{l:\RefL{Tree}\times r:\RefL{Tree}\times\mathsf{rc}:\Uint{32}\times x:\mathsf i_{32}}}}\times\Zero7\times t:\Uint1$
\end{tabular}
\endgroup
\caption{Layouts for $\RefT{Tree}$ in \code{tmap}. Optimized \sys{} is \code{paired-high-tags} discussed in \Cref{sec:dse}.}
\label{fig:evaluation-tmap-layouts}
\end{figure}

The suite is the union of the \ResultMorphicOriginalCount{} benchmarks evaluated in both Morphic papers~\cite{brandon2023better,brandon2026fullyautomatic} in addition to \code{kd-tree}, \code{tree23} (\Cref{sec:evaluation-case-studies}), and \code{tmap} (\Cref{sec:evaluation-allocation}). The benchmark \code{primes-iter} declares no ADTs but remains in the suite for completeness. Following the AST-based translation of Morphic~\cite{brandon2023better}, we generate each MLton and Morphic program from the same \sys{} source, so the comparison measures code generation and data representation rather than how each program was written. Both reference compilers check every array access, so we enable \sys{}'s optional bounds checks.

\sys{}'s arrays are immutable, whereas SML's are mutable. A direct AST translation to MLton therefore cannot safely update an array in place unless no other reference observes the original. For each benchmark, we time four MLton array implementations and use the fastest whose output agrees (Appendix~\ref{app:mlton-arrays}). Morphic uses borrow-based reference counting~\cite{brandon2026fullyautomatic} with a 64-bit count. We acknowledge that cross-language comparison is inherently contrived, and provide these results as a baseline that places \sys{} relative to the state of the art.

\textit{Default layouts}. With default layouts, \sys{} is $\ResultMLtonSpeedup{}\times$ faster than MLton and $\ResultMorphicSpeedup{}\times$ faster than Morphic (geomean, \Cref{fig:evaluation-language-time}). Against MLton, both extremes come from garbage collection, whose copying is costly on the large live sets of \code{rbtree} and \code{rbtree-ck}, but whose bump allocation beats \sys{} on programs with many short-lived objects, e.g., \code{lisp}. Against Morphic, which updates unique arrays in place but never reuses a constructor's cell, dynamic reuse likely contributes to \sys{}'s lead. Appendix~\ref{app:mimalloc-rss} compares peak RSS, which for MLton includes its garbage-collected heap.

\textit{Parity layouts}. To separate layout from code generation, we also run \sys{} at Morphic's own layouts, derived with our rewrite rules and matched bit for bit, in \ResultMatchedLayoutCount{} comparisons (\Aref{app:morphic-layouts}). \sys{} remains $\ResultMatchedMorphicSpeedup{}\times$ faster (geomean), suggesting its lead comes from additional compiler optimizations across instruction selection and memory management.

\begin{figure}[tb]
  \centering
  \captionsetup{skip=2pt}
  \includegraphics[width=\linewidth]{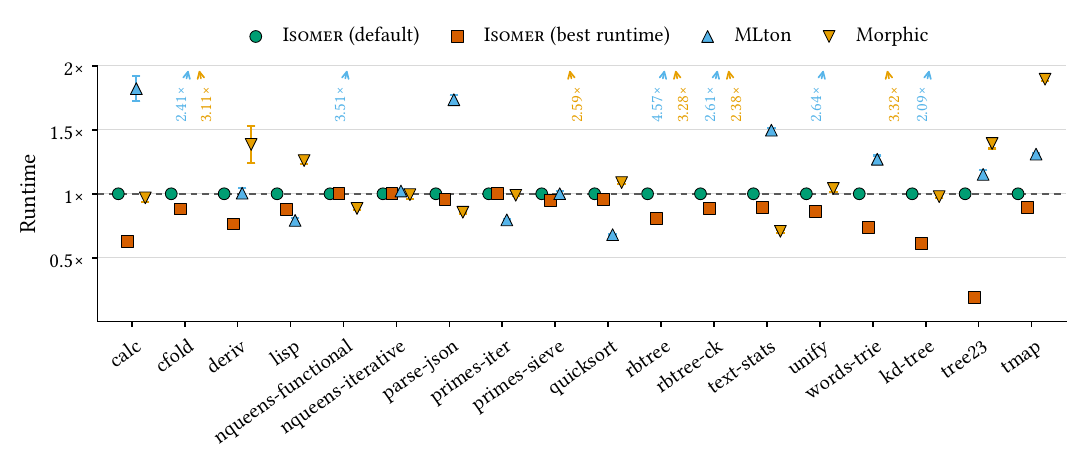}
  \caption{Lower is better. Runtime relative to default \sys{}. The best-runtime layout is selected independently for each benchmark. Arrows indicate ratios beyond the displayed range.}
  \label{fig:evaluation-language-time}
\end{figure}

\subsection{Layout Design Space Exploration}
\label{sec:dse}
We now turn to \emph{exploration} of layouts. Holding each program's source and the compiler fixed, we measure \ResultLayoutCount{} layout configurations across the \ResultBenchmarkCount{} benchmarks of \Cref{sec:evaluation-comparison}. Following Helion's approach of combining language-model proposals with search~\cite{choi2026helion}, we use language models to propose rewrite sequences and further refine the resulting layouts with a straightforward search procedure~\cite{lowerre1976harpy}. Selecting the fastest measured layout for each benchmark gives a $\ResultDSESpeedup{}\times$ geometric-mean speedup over \sys{} with a default layout, ranging from none on \code{primes-iter} (no ADTs) to $\ResultTreeTwoThreeDSESpeedup{}\times$ on \code{tree23}. Independently selecting the lowest-RSS layouts reduces peak RSS by \ResultDSERSSReduction{}\% (geomean). \Cref{fig:evaluation-layouts} shows the frontiers of three benchmarks, and~\Aref{app:layout-frontiers} shows the frontier of every benchmark with a non-default Pareto-optimal layout.

\begin{figure}[tb]
  \centering
  \includegraphics[width=\linewidth]{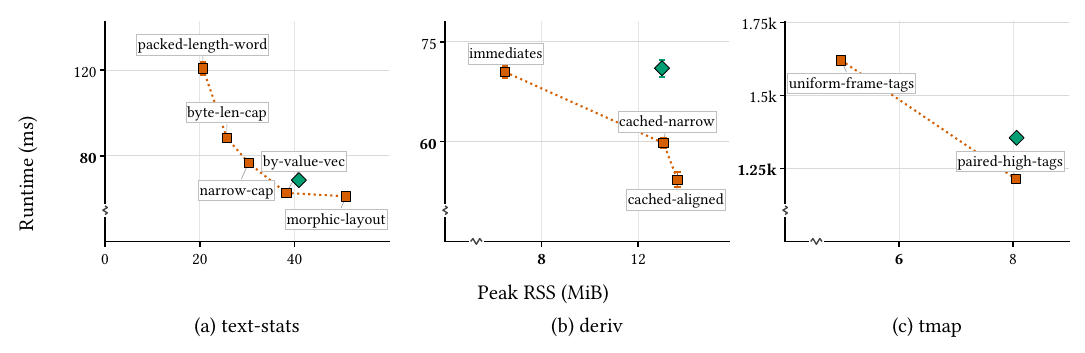}
  \caption{Lower is better. Runtime--RSS tradeoffs among \sys{} layouts for selected benchmarks. Axis break marks indicate nonzero origins. \DSEDefaultMarker{} marks the default \sys{} layout, and \DSEFrontierMarker{} marks a rewritten \sys{} layout on the Pareto frontier.}
  \label{fig:evaluation-layouts}
\end{figure}

The three frontiers of \Cref{fig:evaluation-layouts} come from different mechanisms. In \code{text-stats}, inlining data of vectors avoids an indirection, whereas narrower capacities and lengths save space for short words. Packing lengths into reference words saves further space, but also adds a decoding step when reading. In \code{deriv}, the fast layouts cache each subexpression's leaf count, as the interval tree caches its maximum, so counting reads one field, and \code{immediates} keeps small constants inside the pointer word, which saves a heap cell per constant and reduces peak memory by \ResultDerivImmediateRSSReduction{}\%. Finally, in \code{tmap}, \code{uniform-frame-tags} refines small keys from 32 to 24 bits and matches tree and continuation-frame cell sizes for reuse. The faster \code{paired-high-tags} layout (displayed in \Cref{fig:evaluation-tmap-layouts}) keeps child references adjacent and places the tree tag in the highest bit of the reference word.

\vspace{-0.25em}
\subsection{Comparison with a Data-Description Language}
\label{sec:evaluation-ribbit}

Ribbit~\cite{baudon2024thesis,baudon2023bitstealing} is a layout description language for general-purpose ADTs. We pair Ribbit's fastest measured layout on the three benchmarks of~\citet{baudon2023bitstealing} with the corresponding \sys{} layout, and compile both to LLVM IR. Ribbit has no automatic memory management, while \sys{}'s layouts drop reference counting through the \code{rc}-0 optimization for two benchmarks, and uses reference counting for third. \sys{} achieves a $\ResultRibbitSpeedup{}\times$ geometric-mean speedup (\Cref{fig:evaluation-ribbit}); the gap lies in code generation. For example, Ribbit compiles every function in destination-passing style~\cite[Sec.~5.2]{baudon2024thesis} so every iteration makes an out-of-line call that returns through a pointer, while \sys{} keeps these values in registers. We therefore report the speedup as a comparison of the two implementations, rather than a claim about the performance potential of Ribbit's layout language.

\begin{figure}[tb]
  \centering
  \captionsetup{skip=2pt}
  \begin{subfigure}[t]{0.34\linewidth}
    \vspace{0pt}
    \centering
    \includegraphics[width=\linewidth]{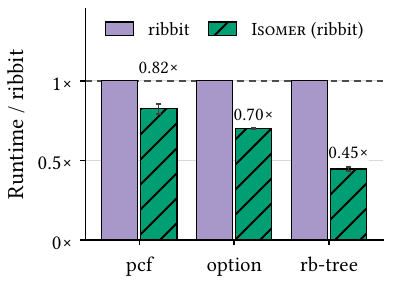}
  \end{subfigure}\hfill
  \begin{subfigure}[t]{0.59\linewidth}
    \vspace{0pt}
    \centering
    \includegraphics[width=\linewidth]{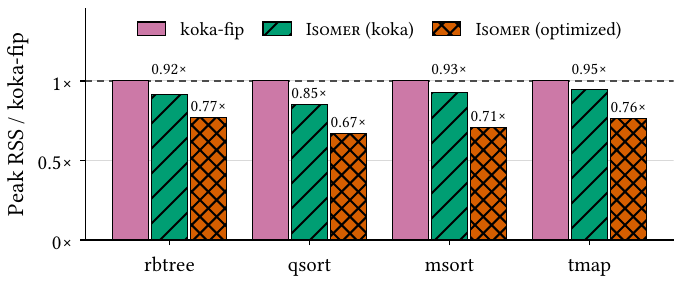}
  \end{subfigure}
  \caption{Lower is better. Runtime relative to Ribbit with corresponding layouts (left), and peak RSS relative to \code{koka-fip} (right). In \sys{} $(\ell)$, $\ell$ denotes the layout used, e.g., \sys{}~(koka) uses Koka's layout.}
  \label{fig:evaluation-ribbit-koka}
  \label{fig:evaluation-ribbit}
  \label{fig:evaluation-allocation}
\end{figure}

\vspace{-0.25em}
\subsection{Allocation and Reuse with Koka}
\label{sec:evaluation-allocation}

We test whether layout control composes with the reference counting optimizations of Koka~\cite{leijen2014koka}, in which a constructor reuses the cell of a same-sized owned value that has just been destructed~\cite{reinking2021perceus}. We run \ResultKokaCount{} of the \ResultKokaSourceCount{} benchmarks of~\citet{lorenzen2023fp2} with $n=\ResultKokaInput{}$, omitting \code{ftree} because it requires polymorphic recursion. We compile Koka with \code{-O3}, and both systems use Koka's allocator. Using Koka's exact layouts, \sys{}'s peak memory is at most that of Koka's fully in-place versions, \code{koka-fip}, on every benchmark (\ResultKokaRSSMatchedReduction{}\% lower peak RSS, geomean; \Cref{fig:evaluation-allocation}). Notably, the fully in-place versions of \code{msort} and \code{qsort} give different constructors the same size by adding \code{Pad} fields to the source~\cite[Sec.~4.3]{lorenzen2023fp2}, whereas \sys{} changes cell sizes solely through layout rewrites. The \code{qsort} program shows the largest gap, which partly is a result of tail recursion modulo cons (TRMC)~\cite{leijen2023trmc}, which \sys{} applies to mutually recursive functions while Koka does not. Further rewrites, such as the bit stealing of \code{tmap} in~\Cref{sec:dse}, lower peak memory by an \emph{additional} \ResultKokaRSSReduction{}\% (geomean), demonstrating that our layout rewrites can complement reuse by reducing the size of allocations that are not optimized away.

\section{Related Work}

\paragraph{Automatic and type-driven layouts} 
MLton~\cite{weeks2006whole} and Morphic~\cite{brandon2023better,brandon2026fullyautomatic} automatically select ADT layouts through whole-program analyses. Gibbon and its successors~\cite{vollmer2017gibbon,vollmer2019local,koparkar2021parallel,singhal2024marmoset} compile recursive ADTs to dense, serialized buffers. Such analyses could select among the layouts our rewrites express. OCaml~\cite{leroy2025ocaml,chataing2024unboxed} and Rust~\cite{ruststdrepr,rustonomiconreprrust} annotations select from a fixed set of layouts without changing application code. Haskell~\cite{peytonjones1991unboxed} and OxCaml~\cite{lorenzen2024oxcaml,janestreet2025oxcaml} expose unboxed types, with levity polymorphism~\cite{eisenberg2017levity} allowing Haskell code to abstract over representation. These choices often require changes to the source code.

\vspace{-0.25em}
\paragraph{Decoupled layouts and schedules} 
Ribbit~\cite{baudon2023bitstealing}, Dargent~\cite{chen2023dargent}, and Scion~\cite{gyurgyik2026decoupling} separate layouts from type declarations but specify the target layout directly. Recently, Rig~\cite{tye2026rig} proposed layout transformations for finite ADTs from type isomorphisms. We instead optimize layouts of recursive ADTs via composable rewrite rules. A layout description language checks each target layout as a whole and decodes it by inverting its final form, so it admits only encodings its checker can validate and its compiler can invert. Our rules carry local proofs and reconstruction functions that compose; a new rule need only satisfy a local correctness criterion rather than modification of a monolithic validation checker. Rewrites also let a programmer or search procedure move between valid layouts incrementally. Dense~\cite{tavakkoli2026lego,zhou2026linearlayouts,lakshminarasimhan2024bricks,hagedorn2020fireiron,shirako2020affinewithlayouts,yadav2022distal} and sparse~\cite{venkat2015sparse,chou2018format,yadav2022spdistal,ahrens2023looplets,liu2024unisparse,liu2026sparseatl} array languages expose layouts for $n$-dimensional arrays. In the same spirit, Halide and related systems separate algorithms from execution schedules for a specific domain~\cite{ragankelley2013halide,chen2018tvm,chen2008chill,senanayake2020scheduling}. \citet{hall2025chronology} identify layout and data movement as further directions. Lift~\cite{steuwer2017lift}, Elevate~\cite{hagedorn2020achieving}, Exo~\cite{ikarashi2022exocompilation,ikarashi2025exo2}, and ATL~\cite{liu2022verified,liu2026sparseatl} compose rewrites for verifiability and extensibility, a principle we apply to data representation.


\vspace{-0.25em}
\paragraph{Data representations and codecs} 
\citet{hoare1989proof} proves correctness through a function from concrete to abstract values, the role of our recovery function. Data description languages generate parsers, printers, and other tools from a format description~\cite{fisher2005pads,bangert2014nail}, and a verified Protocol Buffer compiler proves every serializer and deserializer it generates correct~\cite{ye2019protocol}. EverParse proves its generated parsers correct~\cite{ramananandro2019everparse,swamy2022hardening} and each message's encoding unique~\cite{ramananandro2019everparse}. Narcissus~\cite{delaware2019narcissus} derives verified codecs via local proofs; our destruction policy is likewise assembled from local proofs over structural representations instead of bits. Its formats, like our layouts, may admit several representations per value and must remain injective. These systems verify executable codecs, each for a given format. Similar to our rewrite rules, \citet{geest2017generic} chain verified conversions toward binary encodings, for formats built from fixed-width words, products, and dependent fields.


\bibliographystyle{ACM-Reference-Format}
\bibliography{references}

\appendix

\begingroup
\let\appendixsection\section
\let\appendixsubsection\subsection
\renewcommand{\section}{\FloatBarrier\appendixsection}
\renewcommand{\subsection}{\FloatBarrier\appendixsubsection}

\section{Provenance Lattice}
\label{app:provenance-lattice}

\Cref{fig:provenance-lattice} shows the depth-one provenance lattice of the red-black tree's $\Ctor{Node}$ and $\Ctor{Leaf}$ constructors. Equality of provenances is decidable (\lean{Prov.decidableEquiv}); two provenances read finitely many paths, constructors, and literals, and projecting a value onto these reads preserves both.

\usetikzlibrary{backgrounds}

\newcommand{\provpattern}[2]{%
  \vcenter{\hbox{%
    \begin{tikzpicture}[x=1mm,y=1mm,
      every node/.style={font=\scriptsize,inner sep=0.3pt,outer sep=0pt}]
      \node (root) at (0,1.5) {$N$};
      \node (left) at (-1.7,-1.5) {$#1$};
      \node (right) at (1.7,-1.5) {$#2$};
      \draw[black!70,line width=0.25pt]
        (root.south) -- (left.north)
        (root.south) -- (right.north);
    \end{tikzpicture}%
  }}%
}

\newcommand{\provmatrixstyled}[6]{%
  \begin{tikzpicture}[x=0.48cm,y=0.32cm,baseline=-0.25ex]
    \foreach \x/\y/\admit/\lbl in
      {0/1/#3/LL,1/1/#4/LN,0/0/#5/NL,1/0/#6/NN} {
      \ifnum\admit=1
        \fill[#1] (\x,\y) rectangle ++(1,1);
        \node[font=\scriptsize,inner sep=0pt]
          at (\x+0.5,\y+0.5) {$\lbl$};
      \else
        \node[font=\scriptsize,text=black!30,inner sep=0pt]
          at (\x+0.5,\y+0.5) {$\lbl$};
      \fi
    }
    \draw[#2]
      (0,0) rectangle (2,2) (0,1) -- (2,1) (1,0) -- (1,2);
  \end{tikzpicture}%
}
\newcommand{\provmatrix}[4]{\provmatrixstyled{black!20}{black!55,line width=0.3pt}{#1}{#2}{#3}{#4}}
\newcommand{\provexample}[5]{\provmatrixstyled{#1!35}{#1,line width=0.7pt}{#2}{#3}{#4}{#5}}

\definecolor{provI}{HTML}{0072B2}
\definecolor{provII}{HTML}{E69F00}
\definecolor{provIII}{HTML}{009E73}

\begin{figure}[h]
\captionsetup{skip=3pt}
\centering
\begin{minipage}[t]{0.41\textwidth}
\centering\vspace{0pt}
\begin{tikzpicture}[
  yscale=0.80,
  font=\footnotesize,
  constructor/.style={draw=black!65,circle,fill=white,minimum size=4.5mm,inner sep=1pt},
  example/.style={draw=#1,fill=#1!35,line width=0.7pt},
  subtree/.pic={
    \draw[black!65,fill=white]
      (0,0) -- (-0.17,-0.3) -- (0.17,-0.3) -- cycle;
  }
]
\foreach \offset/\height/\lk/\rk/\col in
  {0/2.8/0/2/provI,3/2.8/1/1/provII,0.4/-0.3/0/2/provIII,2.6/-0.3/1/0/provIII} {
  \begin{scope}[xshift=\offset cm,yshift=\height cm]
    \node[constructor,example=\col] (root) at (0,1.15) {$N$};
    \foreach \side/\kind/\field in {-1/\lk/l,1/\rk/r} {
      \coordinate (child) at (\side*0.65,0.35);
      \draw[black!65] (root) -- (child)
        node[midway,above,font=\scriptsize] {$\field$};
      \ifcase\kind
        \node[constructor,rectangle,example=\col] at (child) {$L$};
      \or
        \node[constructor,example=\col] (childnode) at (child) {$N$};
        \foreach \dx in {-0.28,0.28} {
          \coordinate (grandchild) at ($(child)+(\dx,-0.6)$);
          \draw[black!65] (childnode) -- (grandchild);
          \pic at (grandchild) {subtree};
        }
      \or
        \pic at (child) {subtree};
      \fi
    }
  \end{scope}
}
\foreach \offset/\height/\example/\provname in {
  0/2.8/i/{\provpattern{L}{L}\cup\provpattern{L}{N}},
  3/2.8/ii/{\provpattern{N}{N}},
  1.5/-0.3/iii/{\provpattern{L}{L}\cup\provpattern{L}{N}\cup\provpattern{N}{L}}} {
  \node at (\offset,\height+1.9) {\textbf{(\example)}\quad$\provname$};
}
\node[font=\Large] at (1.5,0.45) {$\cup$};
\end{tikzpicture}

\end{minipage}\hfill
\begin{minipage}[t]{0.57\textwidth}
\centering\vspace{0pt}
\resizebox{\linewidth}{!}{%
\begin{tikzpicture}[
  x=0.95cm,y=0.59cm,
  font=\footnotesize,
  provenance/.style={fill=white,inner sep=2pt},
  cover/.style={draw=black!23,line width=0.5pt},
  lab/.style={fill=white,inner sep=2pt}
]
\node[provenance] (p15) at (0,6.8) {\provmatrix{1}{1}{1}{1}};
\node[provenance] (p7) at (-3,5.1) {\provexample{provIII}{1}{1}{1}{0}};
\node[provenance] (p11) at (-1,5.1) {\provmatrix{1}{1}{0}{1}};
\node[provenance] (p13) at (1,5.1) {\provmatrix{1}{0}{1}{1}};
\node[provenance] (p14) at (3,5.1) {\provmatrix{0}{1}{1}{1}};
\node[provenance] (p3) at (-4,3.4) {\provexample{provI}{1}{1}{0}{0}};
\node[provenance] (p5) at (-2.4,3.4) {\provmatrix{1}{0}{1}{0}};
\node[provenance] (p9) at (-0.8,3.4) {\provmatrix{1}{0}{0}{1}};
\node[provenance] (p6) at (0.8,3.4) {\provmatrix{0}{1}{1}{0}};
\node[provenance] (p10) at (2.4,3.4) {\provmatrix{0}{1}{0}{1}};
\node[provenance] (p12) at (4,3.4) {\provmatrix{0}{0}{1}{1}};
\node[provenance] (p1) at (-3,1.7) {\provmatrix{1}{0}{0}{0}};
\node[provenance] (p2) at (-1,1.7) {\provmatrix{0}{1}{0}{0}};
\node[provenance] (p4) at (1,1.7) {\provmatrix{0}{0}{1}{0}};
\node[provenance] (p8) at (3,1.7) {\provexample{provII}{0}{0}{0}{1}};
\node[provenance] (p0) at (0,0) {\provmatrix{0}{0}{0}{0}};
\begin{scope}[on background layer]
\foreach \lo/\hi in {
  0/1,0/2,0/4,0/8,
  1/3,1/5,1/9,2/3,2/6,2/10,4/5,4/6,4/12,8/9,8/10,8/12,
  3/7,3/11,5/7,5/13,9/11,9/13,6/7,6/14,10/11,10/14,12/13,12/14,
  7/15,11/15,13/15,14/15}
  \draw[cover] (p\lo) -- (p\hi);
\end{scope}
\node[above=2pt] at (p15.north) {$\mathcal{V}_N$};
\node[below=2pt] at (p0.south) {$\bot$};
\node[left=1pt,lab] at (p3.west) {\textbf{(i)}};
\node[right=1pt,lab] at (p8.east) {\textbf{(ii)}};
\node[left=1pt,lab] at (p7.west) {\textbf{(iii)}};
\end{tikzpicture}
}

\end{minipage}
\caption{Constructor provenance of the children $\mathsf l$ and $\mathsf r$ of a $\Ctor{Node}$ of the red-black $\RefT{Tree}$, abbreviating $\Ctor{Leaf}$ as $L$ and $\Ctor{Node}$ as $N$. Triangles denote arbitrary subtrees. Each matrix shades the admitted child pairs, its cell $AB$ standing for children $A$ and $B$, and edges order the matrices by inclusion. The examples, which require (i) a left leaf, (ii) two node children, or (iii) at least one leaf child, share the colors of their matrices.}
\label{fig:provenance-lattice}
\end{figure}

\section{Well-Formedness}
\label{app:well-formedness}

The judgment $A\vdash\ell\;\mathsf{wf}$ checks a layout $\ell$ on its admitted subjects $A$, the source values that can reach $\ell$ (\Cref{sec:layout-provenances}).  A layout definition for $\RefT T$ is verified on all values of its type, $\mathcal V_{\RefT T}$, and each rule checks every immediate sublayout on the subjects that can reach it. The judgment therefore guarantees every admitted subject has at least one representation (\emph{totality}, \Cref{lem:totality}). The rules of \Cref{fig:wf} define the well-formedness judgment of \Cref{sec:well-formedness}, which every rewrite judgment of \Aref{app:rewrite-judgments} requires of both sides.

\begin{figure}[htbp]
\small
\begin{mathpar}
\inferrule{\ell\in\{\epsilon,\Pad n,\Zero n\}}
  {A\vdash\ell\;\mathsf{wf}}
\and
\inferrule{A\subseteq\mathcal V_{\Uint n}}
  {A\vdash\Uint n\;\mathsf{wf}}
\and
\inferrule{A\subseteq\operatorname{dom}\llbracket e\rrbracket}
  {A\vdash\Stored e\;\mathsf{wf}}
\\
\inferrule{A\subseteq\operatorname{dom}\pi \\ A.\pi\vdash\ell\;\mathsf{wf}}
  {A\vdash\pi:\ell\;\mathsf{wf}}
\and
\inferrule{A\vdash\ell\;\mathsf{wf}}
  {A\vdash\Ptr n{\ell}\;\mathsf{wf}}
\and
\inferrule{A\vdash\ell_0\;\mathsf{wf} \\ A\vdash\ell_1\;\mathsf{wf}}
  {A\vdash\ell_0\times\ell_1\;\mathsf{wf}}
\and
\inferrule{\RefT T\in\operatorname{dom}\Delta \\ A\subseteq \mathcal V_{\RefT T}}
  {A\vdash\RefL T\;\mathsf{wf}}
\\
\inferrule
  {A\subseteq\bigcup_j\phi_j \\ A\cap\phi_j\vdash\ell_j\;\mathsf{wf}}
  {A\vdash\Sum{\ar{\phi_j}{\ell_j}}\;\mathsf{wf}}
\and
\inferrule
  {A\subseteq\bigcup_i A_{\pi=c_i}
   \\ A_{\pi=c_i}\subseteq\phi_i
   \\ A_{\pi=c_i}\vdash\ell_i\;\mathsf{wf}}
  {A\vdash\Tagged\pi{\ar{(c_i,\phi_i)}{\ell_i}}\;\mathsf{wf}}
\\
\inferrule{\forall\RefT T\in\operatorname{dom}\Delta,\ \mathcal V_{\RefT T}\vdash\Delta(\RefT T)\;\mathsf{wf}}
  {\Delta\;\mathsf{wf}}
\end{mathpar}
\caption{Well-formedness on the admitted subjects $A$, under the layout environment $\Delta$. We write $\operatorname{dom}\llbracket e\rrbracket$ for the values on which $e$ is defined and $\operatorname{dom}\pi$ for those on which $v.\pi$ is defined. A named layout passes the selected subjects $A.\pi$ to its payload. We write $A_{\pi=c_i}$ for the admitted subjects whose key $\pi$ is defined and equals $c_i$. In the mechanization, a disjoint sum keyed by a tag takes its guards from the tag's declaration, so $A_{\pi=c_i}\subseteq\phi_i$ follows from the selector's soundness. \lean{WF, WF.{sumMap_iff, taggedMap_iff}, LayoutEnv.WF}}
\label{fig:wf}
\end{figure}

\section{Rewrite Judgments}
\label{app:rewrite-judgments}

\Cref{fig:rewrite-judgments} states the rules of \Cref{fig:rewrite-rules} as judgments $A\vdash_\Delta\ell_0\longrightarrow\ell_1$ on admitted subjects $A$ under the layout environment $\Delta$, each establishing local refinement (\Cref{def:local}). A judgment with $\longleftrightarrow$ abbreviates both directions. Besides the displayed premises, each rule requires both sides to be well formed on $A$ (\Aref{app:well-formedness}). Here $j,k,m,n\in\mathbb N$. $\Omega$ and $\Psi$ are arbitrary lists of arms, and an indexed arm $\ar{\phi_j}{\ell_j}$ stands for a list of arms indexed from $0$. In $\mathsf{tag}$, the arms may be listed in any order, and $\mathsf{guards}(t)$ lists their guards in that order (\Cref{sec:layout-language}).

\begingroup
\newcommand{\rwjudg}[2]{A\vdash_\Delta #1\longrightarrow #2}
\newcommand{\rwboth}[2]{A\vdash_\Delta #1\longleftrightarrow #2}
\renewcommand{\RightTirNameStyle}[1]{\scriptsize #1}

\begin{figure}[H]
\footnotesize
\begin{mathpar}
\inferrule*[right=$\mathsf{pad}_n$]{ }{\rwjudg{\ell}{\Pad n\times\ell}}
\and
\inferrule*[right=$\mathsf{zero}_n$]{ }{\rwjudg{\ell}{\Zero n\times\ell}}
\and
\inferrule*[right=$\mathsf{ptr}_n$]{ }{\rwjudg{\ell}{\Ptr n\epsilon\times\ell}}
\and
\inferrule*[right=$\mathsf{memoize}_e$]{ }{\rwjudg{\ell}{\Stored e\times\ell}}
\and
\inferrule*[right=$\mathsf{box}$]{ }{\rwjudg{\Ptr n{\ell_0}\times\ell_1}{\Ptr n{\ell_0\times\ell_1}}}
\and
\inferrule*[right=$\mathsf{unbox}$]{ }{\rwjudg{\Ptr n\ell}{\ell}}
\and
\inferrule*[right=$\mathsf{commute}$]{ }{\rwjudg{\ell_0\times\ell_1}{\ell_1\times\ell_0}}
\and
\inferrule*[right=$\mathsf{assoc}$]{ }{\rwboth{(\ell_0\times\ell_1)\times\ell_2}{\ell_0\times(\ell_1\times\ell_2)}}
\and
\inferrule*[right=$\mathsf{unit}$]{ }{\rwboth{\epsilon\times\ell}{\ell}}
\\
\inferrule*[right=$\mathsf{split}_k$]{k\le n}
  {\rwjudg{\pi:\Uint n}{\pi[0,k):\Uint k\times\pi[k,n):\Uint{n-k}}}
\and
\inferrule*[right=$\mathsf{collapse}$]{ }{\rwjudg{\Sum{\ar\phi\ell}}{\ell}}
\\
\inferrule*[right=$\mathsf{refine}_{\Uint m}$]{ }
  {\rwjudg{\Uint n}{\Sum{\ar{\mathcal V_{\Uint m}}{\Uint m},\ar\top{\Uint n}}}}
\and
\inferrule*[right=$\mathsf{refine}_\phi$]{ }
  {\rwjudg{\ell}{\Sum{\ar\phi\ell,\ar\top\ell}}}
\\
\inferrule*[right=$\mathsf{distribute}$]{ }{\rwjudg{\ell\times\Sum\Omega}{\Sum{\ell\times\Omega}}}
\and
\inferrule*[right=$\mathsf{factor}$]{\Omega\neq\varnothing}{\rwjudg{\Sum{\ell\times\Omega}}{\ell\times\Sum\Omega}}
\\
\inferrule*[right=$\mathsf{prune}$]{A\subseteq\bigcup_j\phi_j}
  {\rwjudg{\Sum{\ar{\phi_j}{\ell_j},\ar\phi\ell}}{\Sum{\ar{\phi_j}{\ell_j}}}}
\and
\inferrule*[right=$\mathsf{flatten}$]{ }
  {\rwjudg{\Sum{\Omega,\ar\phi{\Sum\Psi}}}{\Sum{\Omega,\phi\cap\Psi}}}
\\
\inferrule*[right=$\mathsf{unfold}$]{\Delta\;\mathsf{wf}\ \text{(\Cref{sec:well-formedness})}}{\rwjudg{\RefL T}{\Delta(\RefT T)}}
\and
\inferrule*[right=$\mathsf{tag}$]{\mathsf{guards}(t)=(\phi_j)}
  {\rwjudg{\Sum{\ar{\phi_j}{\ell_j}}}{t:\Uint{|t|}\times\Tagged t{\ar{(j,\phi_j)}{\ell_j}}}}
\end{mathpar}
\caption{Rewrite judgments, named as in \Cref{fig:rewrite-rules}. \lean{Core.Step}}
\label{fig:rewrite-judgments}
\end{figure}
\endgroup

\clearpage
\section{Layout Interpretation}
\label{app:interpretation-example}

Consider $\RefT T=\Ctor A(x:\Uint1,\ y:\Uint1)\mid\Ctor B(y:\Uint1,\ x:\Uint1)$, whose two constructors declare the same one-bit fields but in opposite orders. The derivation below computes the structural representations of $\Ctor B(0,1)$ under a rewritten variant of the canonical layout that stores its field $x$ behind a two-bit pointer. Since this layout refers to no type, we interpret it in the empty environment $\varnothing$. We use $\interpchange{\textbf{blue}}$ to denote an evaluation step.

\begin{figure}[H]
\centering
\begingroup\footnotesize
\[
\begin{aligned}
&\W^{\varnothing}_{\Sum{\ar{\Ctor A}{x:\Uint1\times y:\Uint1},\,\ar{\Ctor B}{y:\Uint1\times\Ptr2{x:\Uint1}}}}\bigl(\Ctor B(0,1)\bigr)\\[0.3ex]
{}={}&\interpchange{\Bigl\{\cons{\Ctor B}\eta\mid\eta\in\W^{\varnothing}_{y:\Uint1\times\Ptr2{x:\Uint1}}\bigl(\Ctor B(0,1)\bigr)\Bigr\}}\\[1.3ex]
&\Bigl\{\cons{\Ctor B}\eta\mid\eta\in\interpchange{\W^{\varnothing}_{y:\Uint1\times\Ptr2{x:\Uint1}}\bigl(\Ctor B(0,1)\bigr)}\Bigr\}\\[0.3ex]
{}={}&\Bigl\{\cons{\Ctor B}\eta\mid\eta\in\interpchange{\W^{\varnothing}_{y:\Uint1}\bigl(\Ctor B(0,1)\bigr)\times\W^{\varnothing}_{\Ptr2{x:\Uint1}}\bigl(\Ctor B(0,1)\bigr)}\Bigr\}\\[1.3ex]
&\Bigl\{\cons{\Ctor B}\eta\mid\eta\in\interpchange{\W^{\varnothing}_{y:\Uint1}\bigl(\Ctor B(0,1)\bigr)}\times\W^{\varnothing}_{\Ptr2{x:\Uint1}}\bigl(\Ctor B(0,1)\bigr)\Bigr\}\\[0.3ex]
{}={}&\Bigl\{\cons{\Ctor B}\eta\mid\eta\in\interpchange{\{0\}}\times\W^{\varnothing}_{\Ptr2{x:\Uint1}}\bigl(\Ctor B(0,1)\bigr)\Bigr\}\\[1.3ex]
&\Bigl\{\cons{\Ctor B}\eta\mid\eta\in\{0\}\times\interpchange{\W^{\varnothing}_{\Ptr2{x:\Uint1}}\bigl(\Ctor B(0,1)\bigr)}\Bigr\}\\[0.3ex]
{}={}&\Bigl\{\cons{\Ctor B}\eta\mid\eta\in\{0\}\times\interpchange{\{\PtrVal{\alpha}{k}\mid \alpha\in\{0,1,2,3\},\ k\in\W^{\varnothing}_{x:\Uint1}\bigl(\Ctor B(0,1)\bigr)\}}\Bigr\}\\[1.3ex]
&\Bigl\{\cons{\Ctor B}\eta\mid\eta\in\{0\}\times\interpchange{\{\PtrVal{\alpha}{k}\mid \alpha\in\{0,1,2,3\},\ k\in\W^{\varnothing}_{x:\Uint1}\bigl(\Ctor B(0,1)\bigr)\}}\Bigr\}\\[0.3ex]
{}={}&\Bigl\{\cons{\Ctor B}\eta\mid\eta\in\{0\}\times\interpchange{\{\PtrVal{0}{1},\,\PtrVal{1}{1},\,\PtrVal{2}{1},\,\PtrVal{3}{1}\}}\Bigr\}\\[1.3ex]
&\interpchange{\Bigl\{\cons{\Ctor B}\eta\mid\eta\in\{0\}\times\{\PtrVal{0}{1},\,\PtrVal{1}{1},\,\PtrVal{2}{1},\,\PtrVal{3}{1}\}\Bigr\}}\\[0.3ex]
{}={}&\interpchange{\left\{\begin{array}{@{}ll@{}}
\cons{\Ctor B}{0,\PtrVal{0}{1}},&
\cons{\Ctor B}{0,\PtrVal{1}{1}},\\
\cons{\Ctor B}{0,\PtrVal{2}{1}},&
\cons{\Ctor B}{0,\PtrVal{3}{1}}
\end{array}\right\}}
\end{aligned}
\]
\endgroup
\caption{The structural representations of $\Ctor B(0,1)$, computed one evaluation step at a time.}
\label{fig:interpretation-example}
\end{figure}

\clearpage
\section{Producer and Consumer Specialization}
\label{app:specialization}
Specialization makes matches read and constructions write the chosen layouts before lowering (\Cref{sec:layout-compilation}). A binder is \emph{inline} when its fields lie directly in the cell, and \emph{stored} when the cell holds it as a scalar or pointer. A construction makes an arm \emph{selectable} when its explicit constructors and literals satisfy the arm's guard. We write $\pi$ for a path and $\kappa$ for the known provenance (\Cref{sec:layout-paths,sec:layout-provenances}).

In the running example, the layout unfolds a list's tail into the head's cell, giving arms $\Ctor{Nil}$, $\Ctor{Cons}(h,\Ctor{Nil})$, and $\Ctor{Cons}(h,\Ctor{Cons}(h',t'))$ with $t'$ behind a pointer. Specialization rewrites \code{map} on the left into the function on the right.
\par\noindent
\begin{minipage}[t]{0.44\linewidth}
\begin{syscode}
fn map(xs: List): List =
  match xs {
    Nil => Nil,
    Cons(h, t) => Cons(h + 1, map(t))
  }
\end{syscode}
\end{minipage}\hfill
\begin{minipage}[t]{0.52\linewidth}
\begin{syscode}
fn map(xs: List): List =
  match xs {
    Nil => Nil,
    Cons(h, Nil) => Cons(h + 1, Nil),
    Cons(h, Cons(h', t')) =>
      Cons(h + 1, Cons(h' + 1, map(t')))
  }
\end{syscode}
\end{minipage}
\par

\paragraph{Consumer specialization.} \Cref{alg:consumer} compiles matches into decision trees against the layout~\cite{maranget2008compiling}. It tests positions demanded by the first live clause in position order, extending $\kappa$ without a test when only one layout arm remains. \textsc{Reify} groups tests within a cell into sibling arms; tests behind a pointer become a nested match. \textsc{Resolve} recompiles matches on inline binders at their paths, inlining one step of a callee that matches the binder. Thus \code{map(t)} becomes \code{Cons(h' + 1, map(t'))}; the call on the stored \code{t'} remains.

\paragraph{Producer specialization.} \Cref{alg:producer} exposes constructors needed to select an arm. \textsc{Expand} inlines a call in a constructor argument and moves the enclosing construction into each resulting match branch. For example, $\Ctor{Cons}(n,\mathit{range}(n-1))$ cannot select an arm until the tail's constructor is known. One step yields $\Ctor{Cons}(n,\Ctor{Nil})$ and $\Ctor{Cons}(n,\Ctor{Cons}(n-1,\mathit{range}(n-2)))$, each selecting an arm. Expansion may continue toward a more specific arm; otherwise, \textsc{Dispatch} tests the child's constructor and respells it in each branch. Stored pointers pass through under aliases recording their constructors. Dispatch requires inline children to decompose into scalars, pointers, or inline constructors that have no further dispatch.

\paragraph{Bounds.} Inlining a recursive call may expose another call without revealing the constructor the layout needs. To ensure compilation terminates, a \emph{budget} bounds this compile-time work, not runtime recursion or input size. Expansion spends one unit per inlined call, starting from $B_\ell=d_\ell(c+1)$. Here $d_\ell$ is the largest number of positions within a value that any one guard tests, excluding the root. In the list layout, the only test beyond the root asks whether the tail is $\Ctor{Nil}$ or $\Ctor{Cons}$, so $d_\ell=1$. The parameter $c$ is the size of the largest group of functions connected by tail calls outside constructions. The factor $c+1$ allows $c$ calls that build nothing plus one that exposes a constructor. Let-bound values restart with their own layout's budget. When expansion runs out, dispatch may expose the missing constructors by runtime tests; it starts from $Q_\ell=d_\ell+1$ and spends one unit per child destruction. If dispatch cannot supply the fields safely, compilation rejects the construction.

Notably, these bounds can reject valid programs. For example, suppose $\mathit{drop}$ traverses a list by tail calls before returning $\Ctor{Nil}$. Expanding $\Ctor{Cons}(h,\mathit{drop}(t))$ exhausts the budget while one branch still contains a recursive call. Under the two-element layout, dispatch cannot safely extract the packed child's fields, so compilation rejects the construction.

\clearpage
\begingroup
\algtext*{EndFunction}
\small
\begin{algorithm}[H]
\caption{Consumer specialization.}
\label{alg:consumer}
\begin{algorithmic}[1]
\Function{SpecializeMatch}{clauses $C$, path $\pi$} \Comment{in the current cell $\ell$, under facts $\kappa$}
  \State \Return \Call{Reify}{the decision tree of $C$ at $\pi$, against $\ell$ under $\kappa$}
\EndFunction
\Statex
\Function{Reify}{$T$}
  \State \textbf{if} $T$ is a leaf $(e,\sigma)$ \textbf{then return} \Call{Resolve}{$e$, $\sigma$} \Comment{$\sigma$: binders to paths}
  \State \Return a match with one reified arm per outcome of $T$
\EndFunction
\Statex
\Function{Resolve}{$e$, $\sigma$}
  \If{$e=\mathsf{match}\ y\ \{\ldots\}$ and $y$ is inline}
    \State \Return \Call{SpecializeMatch}{the clauses of $e$, $\sigma(y)$}
  \ElsIf{$e=g(\ldots,y,\ldots)$, the body of $g$ matches that parameter, and $y$ is inline}
    \State \Return \Call{SpecializeMatch}{the clauses of $g$, $\sigma(y)$} \Comment{inline one step of $g$}
  \ElsIf{$e=y$ and $y$ is stored}
    \State \Return the field or pointer that stores $y$
  \ElsIf{$e=y$ and $y$ is inline as $\Ctor K(a_1,\ldots,a_n)$ with every $a_i$ stored or constant}
    \State \Return $\Ctor K(a_1,\ldots,a_n)$ \Comment{rebuild one constructor}
  \ElsIf{$e=y$}
    \State \textbf{reject} \Comment{rebuilding $y$ would copy its structure}
  \EndIf
  \State \Return $e$ with \textsc{Resolve} applied to each subexpression
\EndFunction
\end{algorithmic}
\end{algorithm}

\begin{algorithm}[H]
\caption{Producer specialization.}
\label{alg:producer}
\begin{algorithmic}[1]
\Function{Expand}{$e$, cell layout $\ell$, budget $b$} \Comment{initially $b=B_\ell$}
  \If{$e=\mathsf{match}\ s\ \{p_i\Rightarrow e_i\}_i$}
    \State \Return $\mathsf{match}\ s\ \{p_i\Rightarrow\Call{Expand}{e_i,\ell,b}\}_i$
  \ElsIf{$e=\mathsf{let}\ x:\ell_x=v\ \mathsf{in}\ e'$}
    \State \Return $\mathsf{let}\ x:\ell_x=\Call{Expand}{v,\ell_x,B_{\ell_x}}\ \mathsf{in}\ \Call{Expand}{e',\ell,b}$
  \ElsIf{$e$ is a construction and some arm of $\ell$ is selectable}
    \State \Return $e$
  \ElsIf{$e=\Ctor K(\ldots,g(\bar u),\ldots)$ for a known function $g$}
    \If{$b>0$}
      \State $e'\gets e$ with $g$ inlined and $\Ctor K$ pushed into the arms of its match
      \State \Return \Call{Expand}{$e'$, $\ell$, $b-1$}
    \ElsIf{\textsc{Dispatch} cannot expose the hidden constructor}
      \State \textbf{reject}
    \EndIf
  \EndIf
  \State \Return $e$
\EndFunction
\Statex
\Function{Dispatch}{$e$, cell layout $\ell$, budget $q$} \Comment{initially $q=Q_\ell$}
  \State \textbf{if} $q=0$, $e$ is not a construction, or an arm is selectable \textbf{then return} $e$
  \State $i\gets$ the first argument whose constructor a guard of $\ell$ constrains
  \ForAll{constructors $\Ctor C$ of $e_i$'s type}
    \State $e_{\Ctor C}\gets\Call{Dispatch}{e[i\leftarrow\Ctor C(\bar y)],\ell,q-1}$ \Comment{$\bar y$ fresh}
  \EndFor
  \State \Return $\mathsf{match}\ e_i\ \{\Ctor C(\bar y)\Rightarrow e_{\Ctor C}\}_{\Ctor C}$
\EndFunction
\end{algorithmic}
\end{algorithm}

\endgroup

\section{Physical Encoding and Decoding}
\label{app:physical-codec}

The encoder and decoder of \Cref{sec:forward-encoder-decoder} satisfy the following sufficient conditions.

\subsection{Discriminant Closure}
\label{app:discriminant-closure}
The judgment $\Gamma\vdash\ell\;\mathsf{disc}$ checks that each tag is stored whole in its scope. $\Gamma$ records available scalar paths, not run-time values. Write $|\pi|$ for a path's declared scalar width. The paths stored whole unconditionally in $\ell$ are (\lean{Stored}):
\[
\begin{aligned}
\mathcal S(\pi:\Uint n)
  &=\begin{cases}\{\pi\},&|\pi|\text{ is defined and at most }n,\\
                    \varnothing,&\text{otherwise},\end{cases}\\
\mathcal S(\Ptr n\ell)&=\mathcal S(\ell),\\
\mathcal S(\ell_0\times\ell_1)&=\mathcal S(\ell_0)\cup\mathcal S(\ell_1),\\
\mathcal S(\ell)&=\varnothing\qquad\text{for all other forms.}
\end{aligned}
\]

At a product $\ell_0\times\ell_1$, check both components with $\Gamma\cup\mathcal S(\ell_0\times\ell_1)$. Pointers preserve $\Gamma$; names reset it to $\varnothing$. Check every sum arm with $\Gamma$, additionally requiring a tagged sum's tag path to belong to $\Gamma$. Leaves need no further check; reference bodies are checked separately from $\varnothing$. The forward variant $\Gamma\vdash\ell\;\mathsf{disc}_{\to}$ instead checks $\ell_0$ with $\Gamma$ and $\ell_1$ with $\Gamma\cup\mathcal S(\ell_0)$, so a tag must precede its payload. Importantly, other decoding orders could relax this requirement when tag locations can be determined without first decoding the payloads they select.

\subsection{Encoding and Decoding Guarantees}
The encoder emits scalars at their widths, most significant bit first, concatenates product components, selects disjoint-sum arms by their tags, and stores each pointer as an address whose heap block contains the pointee. The decoder reads components from left to right. Its sufficient condition is the forward judgment $\varnothing\vdash\ell\;\mathsf{disc}_{\to}$ of \Aref{app:discriminant-closure}, where each tag field must be stored whole before its payload in the same scope. Both functions take a recursion bound $n$ (fuel), written $\mathsf{encode}^n_{\ell}$ and $\mathsf{decode}^n_{\ell}$.

A layout is \emph{concrete} for $\tau$ when the representations of $\mathcal V_\tau$ contain no markers of ordinary sums. For a minimum pointer width $w$, a value $v$ satisfies the \emph{address-capacity bound} when some $\eta_0\in\W^{\Delta}_{\ell}(v)$ has $|\eta_0|_{\mathcal P}\le2^w$, where $|\eta|_{\mathcal P}$ counts the pointer occurrences in $\eta$, so that the encoder can give every occurrence a distinct address.

\begin{theorem}[Encoding within address capacity \lean{Physical.encoding_exists_direct}]
\label[theorem]{thm:encoding-existence}
Suppose $\mathsf{Inv}(\Delta)$, $\ell$ is well formed on $\mathcal V_\tau$ under $\Delta$ and concrete for $\tau$, $\varnothing\vdash\ell\;\mathsf{disc}$, and $\varnothing\vdash\Delta(\RefT T)\;\mathsf{disc}$ for every $\RefT T\in\operatorname{dom}\Delta$. Suppose every pointer in $\ell$ and the bodies bound by $\Delta$ has width at least $w$. For every $v\in\mathcal V_\tau$ satisfying the address-capacity bound, the encoder succeeds at some fuel on some permitted representation,
\[
  \exists\eta\in\W^{\Delta}_{\ell}(v),\ n\in\mathbb N,\ m\in\mathcal M.
  \quad\mathsf{encode}^n_{\ell}(\eta)=m.
\]
\end{theorem}

\begin{lemma}[Recovery by forward decoding \lean{Physical.roundTrip_direct}]
\label[lemma]{lem:storage-round-trip}
If $\Delta$ and $\ell$ are well formed, $\varnothing\vdash\ell\;\mathsf{disc}_{\to}$, and $\varnothing\vdash\Delta(\RefT T)\;\mathsf{disc}_{\to}$ for every $\RefT T\in\operatorname{dom}\Delta$, then at every fuel $n$ the decoder $\mathsf{decode}^n_{\ell}$ returns $\eta$ for every successful encoding $\mathsf{encode}^n_{\ell}(\eta)$ of a permitted representation $\eta$, and so does every larger fuel. Each such pair therefore satisfies the storage recovery property~\Cref{eq:storage-round-trip}.
\end{lemma}

\section{Program Preservation}
\label{app:program-preservation}
Programs consist of first-order function definitions over source values, with the expression syntax (\lean{Programs.{Expr, Prog}})
\[
\begin{array}{@{}r@{\;}c@{\;}l@{\qquad}r@{\;}c@{\;}l@{}}
\multicolumn{6}{c}{x,f,\Ctor K\in\mathsf{Id}\qquad k\in\mathbb N\qquad \mathit{op}:\mathbb N\times\mathbb N\rightharpoonup\mathbb N}\\[0.6ex]
e&::=&x\mid k\mid\mathit{op}(e,e)\mid\Ctor K(\bar e)\mid f(\bar e)
 &P&::=&\{f_i(\bar x_i)=e_i\}.\\
 &\mid&\mathbf{let}\ x=e\ \mathbf{in}\ e\mid\mathbf{match}\ e\ \mathbf{with}\ \{\Ctor K_i(\bar x_i)\Rightarrow e_i\}
\end{array}
\]
For simplicity, function and match-arm lookup select the first matching name. Both interpretations evaluate arguments from left to right before construction or a call. A call requires the declared arity and evaluates the function body in a fresh environment containing only its parameter bindings; recursive calls use the same $P$. Variables, literals, scalar operations, and let bindings have their usual call-by-value meaning. An undefined lookup or scalar operation supplies no evaluation derivation.

Source evaluation $P;\rho\vdash e\Downarrow v$ uses the values of \Cref{sec:layout-paths}. Construction evaluates its arguments to $\bar v$ and forms $\Ctor K(\bar v)$, provided this value belongs to the constructor's declared type. Matching selects the arm for the evaluated constructor and binds its fields to the branch variables in declaration order, with equal arities. Structural evaluation uses the same program and scalar operations, but uses represented datatype values, defined below.

\paragraph{The datatype interface}
Represented values comprise scalars and datatype handles. For a relation $r\sim v$ from represented values to source values, the interface provides partial operations
\[
  \mathsf{construct}(\Ctor K,\bar r),
  \qquad
  \mathsf{destruct}(r).
\]
Scalars relate only to the same scalar, datatype handles relate only to constructor values, and each represented value relates to at most one source value. For componentwise-related arguments $\bar r\sim\bar v$, construction succeeds exactly when $\Ctor K(\bar v)$ is admitted by the declaration, including its field types and arity; its result relates to $\Ctor K(\bar v)$. Destruction of a handle related to $\Ctor K(\bar v)$ succeeds with $(\Ctor K,\bar r)$ where $\bar r\sim\bar v$ in declaration order. These conditions give both directions of the constructor and match cases in evaluation correspondence.

The remainder instantiates this interface with faithful layouts, through a construction policy $b$ and a destruction policy $\iota$ satisfying
\[
  b_{\RefT T}(v)\in\W^\Delta_{\RefL T}(v),
  \qquad
  \iota_{\RefT T}(\eta)=v
  \quad\text{for }v\in\mathcal V_{\RefT T}\text{ and }\eta\in\W^\Delta_{\RefL T}(v).
\]
A valid $b$ meets the left condition on every admitted value and returns nothing for any other value, and a valid $\iota$ meets the right one. 

\paragraph{Computable policies} 
The encoder of \Cref{lem:totality} gives a computable valid $b$ under $\mathsf{Inv}(\Delta)$ when $\Delta$ binds every declared type, and the canonical decoder, composed with the reconstruction functions of further rewrites, gives a valid $\iota$. These are computable functions of the model; implementing them as generated code requires a separate proof of the same interface laws.

\begin{theorem}[Derived policies \lean{Core.UpdateSeq.{decoder_decodes,observations}}]
\label[theorem]{thm:derived-policies}
Let $\Delta_0$ be a canonical environment and $\Delta_0\longrightarrow^*\Delta_1$ a finite sequence of valid updates that keep their reconstruction functions. Undoing the updates in reverse order, each by applying its function at every reference to the updated type and then applying the canonical decoder gives a valid destruction policy $\iota_1$ for $\Delta_1$. Hence \eqref{eq:program-observations} holds with the encoder of \Cref{lem:totality} at both endpoints, the canonical decoder at $\Delta_0$, and $\iota_1$ at $\Delta_1$, for all related inputs.
\end{theorem}

\paragraph{Views and related inputs}
A represented value has the form
\[
  r ::= k\mid\langle\RefT T,\eta,\pi\rangle.
\]
The type $\RefT T$ describes the value represented by $\eta$; the path $\pi$ selects a datatype subvalue of that value, possibly of another type. Scalars represent themselves. A view represents a constructor value $w$ exactly when an admitted root has that representation and selects $w$:
\[
\begin{aligned}
  k\sim_\Delta v
    &\iff v=k,\\
  \langle\RefT T,\eta,\pi\rangle\sim_\Delta w
    &\iff w\text{ is a constructor value and}\\
    &\quad\exists u\in\mathcal V_{\RefT T}.\quad
      \eta\in\W^\Delta_{\RefL T}(u)\ \land\ u.\pi=w.
\end{aligned}
\]
Environments $\rho'\sim_\Delta\rho$ have the same domain and satisfy $\rho'(x)\sim_\Delta\rho(x)$ at each bound variable. Thus related inputs provide both evaluations with the same source values, while their structural representations may differ.

A destruction policy $\iota$ induces the partial reader
\[
  \alpha_\iota(k)=k,
  \qquad
  \alpha_\iota(\langle\RefT T,\eta,\pi\rangle)
    =\bigl(\iota_{\RefT T}(\eta)\bigr).\pi.
\]
Under a valid $\iota$, related values satisfy $r\sim_\Delta v\implies\alpha_\iota(r)=v$. Structural construction first evaluates its arguments to $\bar r$ and reads $\bar v=\alpha_\iota(\bar r)$ componentwise. For $v=\Ctor K(\bar v)\in\mathcal V_{\RefT T}$, the policy $b$ returns a permitted representation and construction produces
\[
  \langle\RefT T,b_{\RefT T}(v),\epsilon\rangle.
\]
The policies are fixed for an evaluation. We represent each as a partial function on arbitrary type names and values or representations; validity of $b$ requires a permitted result for every value of each declared type and no result for any other value. The functions $b_{\RefT T}$ above are its total restrictions to those values.

Matching evaluates its scrutinee to $\langle\RefT T,\eta,\pi\rangle$ and obtains its constructor through $\alpha_\iota$. For declared fields $f_i:\tau_i$, it binds scalar fields to their recovered scalar values and datatype fields to $\langle\RefT T,\eta,\pi.f_i\rangle$. It then evaluates the selected branch in the extended environment. The extended views keep $\eta$ and need no separately constructed representation of a field. The branch variables must match the declared field count. These operations and the common rules define $\Delta,b,\iota;P;\rho'\vdash e\Downarrow r$.

\begin{lemma}[Evaluation correspondence \lean{Programs.evaluation_value_iff}]
\label[lemma]{lem:evaluation-correspondence}
Suppose $\Delta$ is faithful. For valid policies $b$ and $\iota$ and related environments $\rho'\sim_\Delta\rho$,
\[
  P;\rho\vdash e\Downarrow v
  \quad\Longleftrightarrow\quad
  \exists r.\quad
    \Delta,b,\iota;P;\rho'\vdash e\Downarrow r
    \ \land\ r\sim_\Delta v.
\]
\end{lemma}

Applying correspondence to both environments gives~\Cref{thm:representation-independence}. Preserving faithfulness of the environment then gives~\Cref{thm:all-rule-program-preservation}. A view of a field continues to refer to its enclosing representation, so this argument requires no agreement between source match depth and layout unfolding depth.

\clearpage
\section{Additional Evaluation Results}
\label{app:evaluation}

\subsection{Complete Layout Frontiers}
\label{app:layout-frontiers}

\DeclareRobustCommand{\DSEMLtonMarker}{%
  \tikz[baseline=-0.4ex,x=1em,y=1em]{%
    \filldraw[fill={rgb,255:red,86;green,180;blue,233},draw=black,line width=0.4pt]
      (-0.3,-0.2) -- (0.3,-0.2) -- (0,0.4) -- cycle;
  }%
}
\DeclareRobustCommand{\DSEMorphicMarker}{%
  \tikz[baseline=-0.4ex,x=1em,y=1em]{%
    \filldraw[fill={rgb,255:red,230;green,159;blue,0},draw=black,line width=0.4pt]
      (-0.3,0.3) -- (0.3,0.3) -- (0,-0.3) -- cycle;
  }%
}
\DeclareRobustCommand{\DSEFrontierSquare}{%
  \tikz[baseline=-0.4ex,x=1em,y=1em]{%
    \filldraw[fill=DSEFrontierColor,draw=black,line width=0.35pt]
      (-0.2,-0.1) rectangle (0.2,0.3);
  }%
}
\DeclareRobustCommand{\DSEFrontierLine}{%
  \tikz[baseline=-0.4ex,x=1em,y=1em]{%
    \draw[DSEFrontierColor,densely dotted,line width=0.8pt] (-0.7,0) -- (0.7,0);
  }%
}

\begin{figure}[H]
  \centering
  \includegraphics[width=\linewidth,height=0.8\textheight,keepaspectratio]{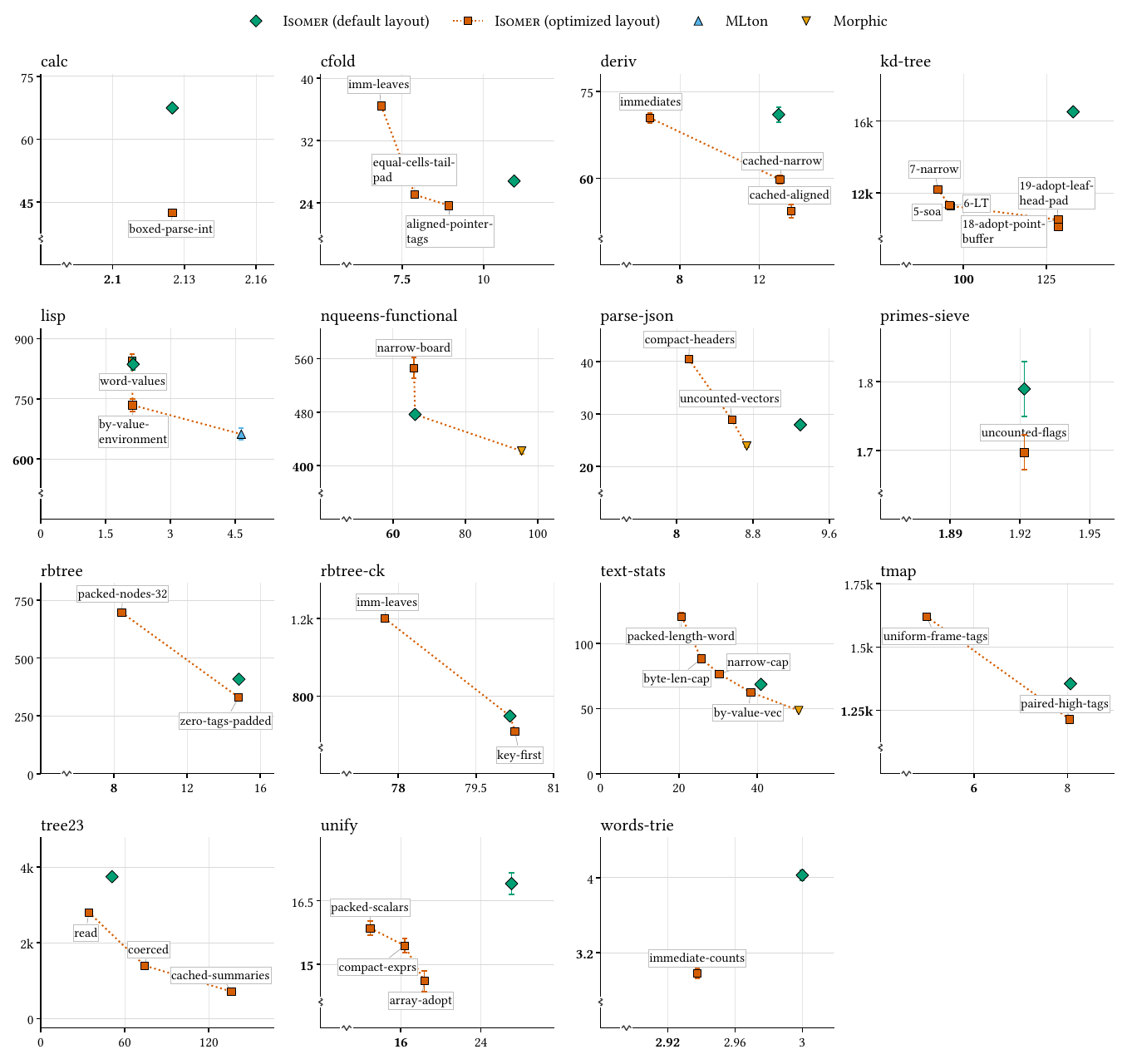}
  \caption{Lower is better. Mimalloc runtime--RSS frontiers for benchmarks with a non-default Pareto-optimal layout. Each panel plots runtime (ms, vertical) against peak RSS (MiB, horizontal). \DSEDefaultMarker{} marks a default \sys{} layout; \DSEFrontierMarker{} marks an \sys{} layout optimized via rewrites on the frontier (\DSEFrontierSquare{}) and traces the frontier (\DSEFrontierLine{}). Triangles mark MLton (\DSEMLtonMarker{}) and Morphic (\DSEMorphicMarker{}). Dominated layouts are omitted.}
  \label{fig:layout-frontiers}
\end{figure}

\subsection{Peak RSS (Mimalloc)}
\label{app:mimalloc-rss}

We report whole-process peak resident set size (RSS), normalized to \sys{} with its default layout. For MLton, this measure includes its GC-managed heap.

\begin{figure}[H]
  \centering
  \includegraphics[width=\linewidth]{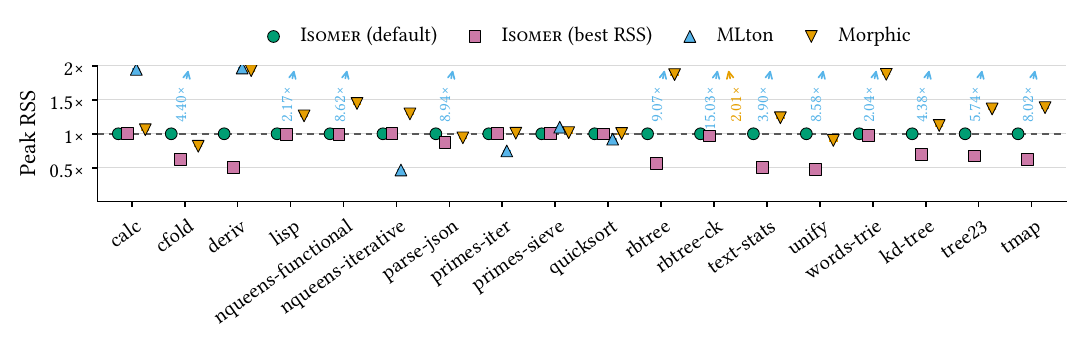}
  \caption{Lower is better. Mimalloc peak RSS relative to default \sys{}. The best-RSS layout is selected independently for each benchmark. Arrows indicate ratios beyond the displayed range.}
  \label{fig:evaluation-language-rss-mimalloc}
\end{figure}

\subsection{Parity Layouts with Morphic}
\label{app:morphic-layouts}

We run \sys{} with Morphic's own layouts, derived with our rewrite rules and matched bit for bit, so that only the compilers differ. One program, \code{primes-iter}, declares no ADTs. \sys{} is $\ResultMatchedMorphicSpeedup{}\times$ faster (geomean).

\begin{figure}[H]
  \centering
  \includegraphics[width=\linewidth]{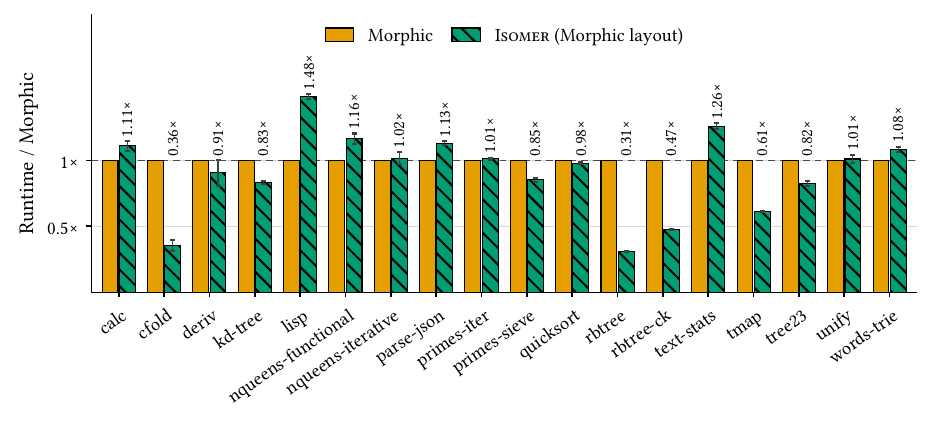}
  \caption{Lower is better. Mimalloc runtime relative to Morphic, with \sys{} at Morphic's layouts.}
  \label{fig:evaluation-morphic-layout}
\end{figure}

\FloatBarrier
\noindent\begin{minipage}{\linewidth}
\subsection{System Allocator Results}
\label{app:system-results}

We repeat the workloads and sampling protocol of \Cref{sec:methodology} with the system allocator. MLton uses its own tracing heap, so its measurements are shared across allocator comparisons. \sys{} with default layout achieves speedups (geomean) of $\ResultSystemMLtonSpeedup{}\times$ versus MLton and $\ResultSystemMorphicSpeedup{}\times$ versus Morphic, using $\ResultSystemMLtonRSSRatio{}\times$ and $\ResultSystemMorphicRSSRatio{}\times$ their peak RSS, respectively. Morphic-style layouts give a $\ResultSystemMatchedMorphicSpeedup{}\times$ speedup over Morphic across all \ResultMatchedLayoutCount{} benchmarks. Against Ribbit on \code{pcf}, \code{option}, and \code{rb-tree}, \sys{} achieves speedups of $\ResultSystemRibbitPCFSpeedup{}\times$, $\ResultSystemRibbitOptionSpeedup{}\times$, and $\ResultSystemRibbitRBSpeedup{}\times$, respectively, or $\ResultSystemRibbitSpeedup{}\times$ (geomean).

\begin{figure}[H]
  \centering
  \includegraphics[width=\linewidth]{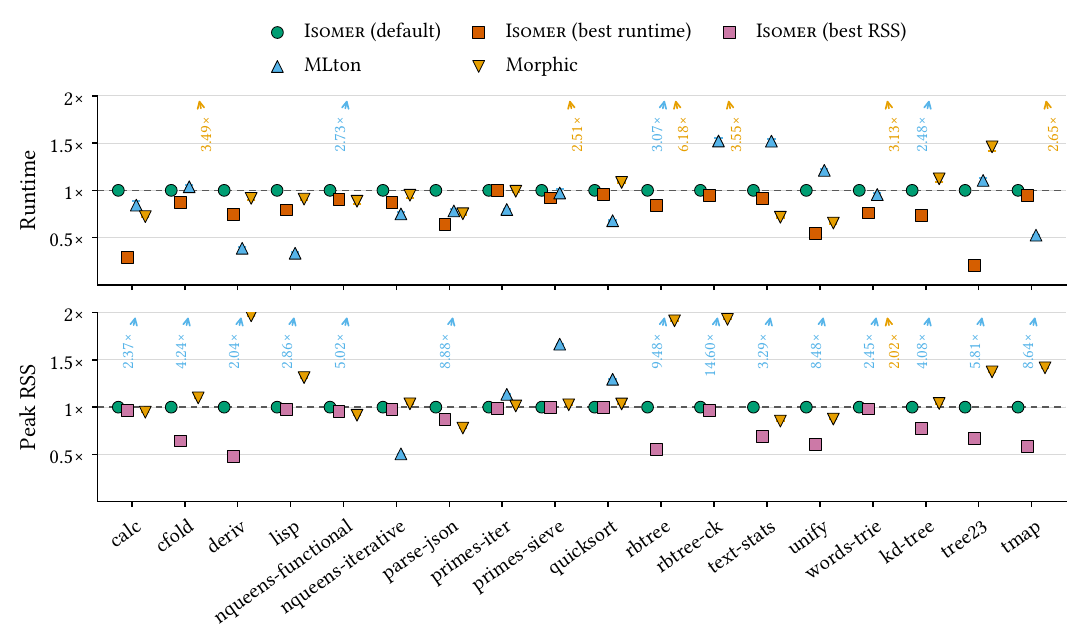}
  \caption{Lower is better. System-allocator runtime (top) and peak RSS (bottom) relative to default \sys{}, using the conventions of \Cref{fig:evaluation-language-time}. The best-runtime and best-RSS layouts are selected independently for each benchmark. Arrows indicate ratios beyond the displayed range.}
  \label{fig:evaluation-language-system}
  \label{fig:evaluation-language-rss-system}
\end{figure}
\end{minipage}

\subsection{MLton Array Representations}
\label{app:mlton-arrays}

For each benchmark, calibration selects the fastest of four MLton array implementations whose output agrees with the reference on the checked workload. The selected implementation is recorded and used for every timed run. \code{primes-iter} and \code{tmap} contain no source arrays.

\begin{table}[htbp]
  \centering
  \caption{MLton array implementations selected for the comparison benchmarks.}
  \label{tab:evaluation-mlton-arrays}
  \begin{tabular}{@{}>{\raggedright\arraybackslash}p{0.13\linewidth}
    >{\raggedright\arraybackslash}p{0.29\linewidth}
    >{\raggedright\arraybackslash}p{0.53\linewidth}@{}}
    \toprule
    Mode & Array Implementation & Benchmarks \\
    \midrule
    \code{mutate} & mutable &
    \code{calc}, \code{nqueens-iterative}, \code{parse-json},
    \code{primes-sieve}, \code{quicksort}, \code{rbtree-ck},
    \code{text-stats}, \code{words-trie}, \code{kd-tree}, \code{tmap} \\
    \code{copy} & Flat, copied on update &
    \code{lisp}, \code{rbtree}, \code{tree23} \\
    \code{reroot} & Persistent, update chains &
    \code{nqueens-functional}, \code{primes-iter}, \code{unify} \\
    \code{trie32} & Persistent, 32-way trie &
    \code{cfold}, \code{deriv} \\
    \bottomrule
  \end{tabular}
\end{table}

\section{Mechanization}
\label{app:mechanization}
\Cref{tab:mechanization} lists, in paper order, every definition, lemma, and theorem of \Cref{sec:layout-language,sec:representation-semantics,sec:rewrite-correctness,sec:physical-storage} and of the appendices, together with the mechanized examples of \Cref{sec:motivation,sec:core-language}. Files are relative to the root of the Lean development, with the prefix \code{Ismr/} omitted. We additionally provide \code{Paper.lean}, which restates each numbered result. Propositional extensionality and quotient soundness are the only axioms used in this formalization. 

\begingroup
\fontsize{7}{7.5}\selectfont
\setlength{\tabcolsep}{3pt}
\setlength{\defaultaddspace}{1.5pt}
\renewcommand{\leanname}[1]{{\normalfont\urlstyle{tt}\def\UrlBreaks{\do\.\do\,\do\{\do\}\do\_}\def\UrlBigBreaks{}\expandafter\nolinkurl\expandafter{\detokenize{#1}}}}
\def\leanfilesplit#1/#2\leanfileend{\def\leanfilehead{#1}\def\leanfiletail{#2}}
\def\leanfileismr{Ismr}
\renewcommand{\leanfile}[1]{{\leanfilesplit#1\leanfileend\ifx\leanfilehead\leanfileismr\edef\leanfileshown{\leanfiletail}\else\edef\leanfileshown{#1}\fi\urlstyle{tt}\def\UrlBreaks{\do\/}\def\UrlBigBreaks{}\expandafter\nolinkurl\expandafter{\leanfileshown}}}
\begin{longtable}{@{}>{\raggedright\arraybackslash}p{\dimexpr0.25\linewidth-2\tabcolsep\relax}>{\raggedright\arraybackslash}p{\dimexpr0.29\linewidth-2\tabcolsep\relax}>{\raggedright\arraybackslash}p{0.46\linewidth}@{}}
\caption{The definitions, theorems, and claims of the paper and its appendices.}
\label{tab:mechanization}\\
\toprule
Paper & File & Lean \\
\midrule
\endfirsthead
\toprule
Paper & File & Lean \\
\midrule
\endhead
\multicolumn{3}{@{}l}{\textit{\Cref{sec:motivation}}}\\
\Cref{lst:morphic,lst:low-bits,lst:shape,lst:terminal}, derived by the rules and faithful & \leanfile{Examples/Motivation.lean} & \leanin{Core}{Examples.Motivation.{morphic_core, lowBits_core, shape_core, terminal_core, morphic_stores, morphic_faithful, lowBits_faithful, shape_faithful, terminal_faithful}}\\
\addlinespace
\multicolumn{3}{@{}l}{\textit{\Cref{sec:layout-language}}}\\
\Cref{sec:layout-paths}, values, typing, and paths & \leanfile{Ismr/Source.lean} & \leanin{Core}{Value, Values, Value.select, Decls.WF}\\
\Cref{sec:layout-provenances}, provenances as sets of values, admitted subjects & \leanfile{Ismr/Source.lean} & \leanin{Core}{Prov.holds}\\
 & \leanfile{Ismr/Provenance.lean} & \leanin{Core}{Prov.Equiv}\\
 & \leanfile{Ismr/Layout/Basic.lean} & \leanin{Core}{Domain}\\
\Cref{fig:layout-grammar}, layouts & \leanfile{Ismr/Layout/Basic.lean} & \leanin{Core}{Layout, LayoutEnv}\\
 & \leanfile{Ismr/Source.lean} & \leanin{Core}{Path, Prov}\\
 & \leanfile{Ismr/Expr.lean} & \leanin{Core}{Expr}\\
Arms and their guards & \leanfile{Ismr/Layout/SumArm.lean} & \leanin{Core}{SumArm, Prov.Unique}\\
 & \leanfile{Ismr/Layout/ArmMaps.lean} & \leanin{Core}{ArmMap}\\
Tags, selectors, and code widths & \leanfile{Ismr/Source.lean} & \leanin{Core}{Decls.lookupTag, ArmSelector, ArmSelector.{sound, complete}, codeWidth, Decls.tagWidth}\\
Canonical layout environment & \leanfile{Ismr/Layout/Canonical.lean} & \leanin{Core}{Canonical.env}\\
\Cref{sec:layout-rewrites}, layout contexts & \leanfile{Ismr/Layout/Context.lean} & \leanin{Core}{Ctxt, Ctxt.{fill, holeDom}}\\
\Cref{fig:rewrite-rules}, rules and certificates & \leanfile{Ismr/Core/Rules.lean} & \leanin{Core}{Step, Step.{certificate, local}}\\
 & \leanfile{Ismr/Core/Local.lean} & \leanin{Core}{RuleCertificate}\\
 & \leanfile{Ismr/Layout/Flatten.lean} & \leanin{Core}{refineArms, flattenArms, WF.flatten, Rule.flattenRecon, Rule.flattenRecon_rep}\\
 & \leanfile{Ismr/Provenance/KeepFirst.lean} & \leanin{Core}{Prov.keepFirst, Prov.{keepFirst_pairwise, keepFirst_covers, find?_guard_of_mem_keepFirst}}\\
Definition updates & \leanfile{Ismr/Layout/Basic.lean} & \leanin{Core}{LayoutEnv.set}\\
\Cref{fig:rewrite-example}, the double-ended layout & \leanfile{Examples/Motivation.lean} & \leanin{Core}{Examples.Motivation.{doubleEnded_core, doubleEnded_stores, doubleEnded_faithful}}\\
\addlinespace
\multicolumn{3}{@{}l}{\textit{\Cref{sec:representation-semantics}}}\\
\Cref{sec:structural-values}, structural representations & \leanfile{Ismr/Layout/Semantics.lean} & \leanin{Core}{Enc}\\
\Cref{sec:structural-values}, the canonical representation of the example node & \leanfile{Examples/Worked.lean} & \leanin{Core}{Examples.Worked.{canonical_node, canonical_node_only}}\\
\Cref{fig:layout-interpretation}, interpretation & \leanfile{Ismr/Layout/Semantics.lean} & \leanin{Core}{Rep}\\
\Cref{def:faithful}, faithfulness and determination & \leanfile{Ismr/Correctness/Faithfulness.lean} & \leanin{Core}{Faithful, Total, AbsoluteInjective, Determines, AbsoluteInjective.determines}\\
\Cref{sec:well-formedness} and \Cref{fig:wf}, well-formedness & \leanfile{Ismr/Layout/WellFormed.lean} & \leanin{Core}{WF, LayoutEnv.WF}\\
 & \leanfile{Ismr/Layout/ArmMaps.lean} & \leanin{Core}{WF.sumMap_iff}\\
 & \leanfile{Ismr/Layout/TaggedArms.lean} & \leanin{Core}{WF.taggedMap_iff}\\
\Cref{def:rewrite-invariant}, environment invariant & \leanfile{Ismr/Core/Local.lean} & \leanin{Core}{Inv}\\
 & \leanfile{Ismr/Layout/WellFormed.lean} & \leanin{Core}{LayoutEnv.WF}\\
 & \leanfile{Ismr/Layout/Structural.lean} & \leanin{Core}{LayoutEnv.Descent}\\
\Cref{lem:totality}, totality by a computable encoder & \leanfile{Ismr/Correctness/Totality.lean} & \leanin{Core}{WF.totality, represent_total}\\
 & \leanfile{Ismr/Layout/Represent.lean} & \leanin{Core}{represent, represent_sound}\\
\addlinespace
\multicolumn{3}{@{}l}{\textit{\Cref{sec:rewrite-correctness}}}\\
\Cref{sec:local-reconstruction}, sharing, refinement, and its laws & \leanfile{Ismr/Core/Local.lean} & \leanin{Core}{Shares, absoluteInjective_iff_shares, Refines, Refines.{refl, trans, faithful}}\\
\Cref{def:local}, local refinement & \leanfile{Ismr/Core/Local.lean} & \leanin{Core}{LocalRefinement}\\
\Cref{sec:local-reconstruction}, reconstruction functions & \leanfile{Ismr/Core/Local.lean} & \leanin{Core}{Reconstructs, Reconstructs.refines, LocalRefinement.of_reconstructs}\\
 & \leanfile{Ismr/Core/Rules.lean} & \leanin{Core}{Step.certificate}\\
 & \leanfile{Ismr/Core/StorageRules.lean} & \leanin{Core}{Recon.{dropFirst, swap, join}}\\
\Cref{thm:contextual-replacement}, contextual replacement & \leanfile{Ismr/Core/Contextual.lean} & \leanin{Core}{Refines.context, LocalRefinement.context}\\
\Cref{lem:recursive}, recursive binding & \leanfile{Ismr/Core/Recursive.lean} & \leanin{Core}{LocalRefinement.bind, Refines.bind}\\
\Cref{thm:rewrite-soundness}, valid rewrites and soundness & \leanfile{Ismr/Core/Correctness.lean} & \leanin{Core}{Rewrite, RewriteSeq, Rewrite.sound, RewriteSeq.sound}\\
\Cref{sec:faithfulness-under-rewriting}, composition and canonical starts & \leanfile{Ismr/Core/Correctness.lean} & \leanin{Core}{RewriteSeq.faithful}\\
 & \leanfile{Ismr/Layout/Canonical.lean} & \leanin{Core}{Canonical.body_wf}\\
 & \leanfile{Ismr/Layout/CanonicalDecode.lean} & \leanin{Core}{Canonical.{faithful, decode_rep}}\\
 & \leanfile{Ismr/Core/Canonical.lean} & \leanin{Core}{canonical_rewrites_faithful}\\
\Cref{sec:extending-rewrites}, interval encoding & \leanfile{Ismr/Core/Interval.lean} & \leanin{Core}{Interval.{Bounds, certificate, local_refinement, rewrite}}\\
\Cref{sec:program-preservation}, the two evaluations, valid policies, related values & \leanfile{Ismr/Programs/SourceSemantics.lean} & \leanin{Core}{Programs.Eval}\\
 & \leanfile{Ismr/Programs/StructuralSemantics.lean} & \leanin{Core}{Programs.StructuralEval}\\
 & \leanfile{Ismr/Programs/Views.lean} & \leanin{Core}{Programs.{Builder, Decoder, Builds, Decodes, Represents, FaithfulRoots}}\\
 & \leanfile{Ismr/Correctness/ProgramPreservation.lean} & \leanin{Core}{Programs.EnvRel}\\
\Cref{thm:representation-independence}, representation independence & \leanfile{Ismr/Correctness/ProgramPreservation.lean} & \leanin{Core}{Programs.representation_independence}\\
 & \leanfile{Ismr/Programs/ViewInterface.lean} & \leanin{Core}{Programs.view_laws}\\
 & \leanfile{Ismr/Programs/Interface.lean} & \leanin{Core}{Programs.Interface.independence}\\
\Cref{thm:all-rule-program-preservation}, program preservation & \leanfile{Ismr/Core/Programs.lean} & \leanin{Core}{BodyRewriteSeq.{observations, roots}}\\
 & \leanfile{Ismr/Layout/Structural.lean} & \leanin{Core}{LayoutEnv.Binds}\\
\Cref{sec:program-preservation}, computable policies & \leanfile{Ismr/Programs/Construct.lean} & \leanin{Core}{Programs.Builds.construct}\\
 & \leanfile{Ismr/Programs/CanonicalDecoder.lean} & \leanin{Core}{Programs.Decodes.canonical}\\
 & \leanfile{Ismr/Core/Policies.lean} & \leanin{Core}{UpdateSeq.{decoder, decoder_decodes}}\\
\addlinespace
\multicolumn{3}{@{}l}{\textit{\Cref{sec:physical-storage}}}\\
\Cref{eq:storage-round-trip}, storage recovery & \leanfile{Ismr/Physical/Semantic.lean} & \leanin{Core}{Physical.{SemanticRoundTrip, RoundTrip}, Physical.RoundTrip.semantic}\\
\Cref{thm:storage-recovery}, recovery, loaded inputs & \leanfile{Ismr/Core/Storage.lean} & \leanin{Core}{storage_recovery}\\
 & \leanfile{Ismr/Physical/Semantic.lean} & \leanin{Core}{Physical.SemanticRoundTrip.recovery}\\
 & \leanfile{Ismr/Programs/SemanticStorage.lean} & \leanin{Core}{Programs.loaded_evaluation_iff}\\
\Cref{sec:forward-encoder-decoder}, the forward codec and a stored node & \leanfile{Ismr/Physical/Coding.lean} & \leanin{Core}{Physical.{encodeLayout, decodeLayout}}\\
 & \leanfile{Ismr/Physical/Direct.lean} & \leanin{Core}{Physical.{encoding_exists_direct, roundTrip_direct}}\\
 & \leanfile{Examples/Motivation.lean} & \leanin{Core}{Examples.Motivation.{lowBits_encode, lowBits_round_trip, lowBits_recovery}}\\
\Cref{sec:decoding-limitations}, read order, ambiguity & \leanfile{Ismr/Physical/Counterexamples.lean} & \leanin{Core}{Physical.Limits.{read_order, encoding_ambiguity, no_value_decoder}}\\
 & \leanfile{Ismr/Physical/Semantic.lean} & \leanin{Core}{Physical.SemanticRoundTrip.injective}\\
\addlinespace
\multicolumn{3}{@{}l}{\textit{\Cref{sec:implementation}}}\\
\Cref{sec:core-language}, a finite array as a sum indexed by its length & \leanfile{Examples/Array.lean} & \leanin{Core}{Examples.Array.{arms, sized, tagSize, updates, sized_faithful, construct_a79, round_trip}}\\
\addlinespace
\multicolumn{3}{@{}l}{\textit{\Aref{app:provenance-lattice}}}\\
\Aref{app:provenance-lattice}, deciding provenance equality & \leanfile{Ismr/Provenance/Decide.lean} & \leanin{Core}{Prov.{decidableEquiv, equivCheck_iff, counterexample, counterexample_sound, counterexample_complete}}\\
 & \leanfile{Ismr/Provenance/Finite.lean} & \leanin{Core}{Observation.{proj, proj_mem, holds_proj}}\\
\addlinespace
\multicolumn{3}{@{}l}{\textit{\Aref{app:rewrite-judgments}}}\\
\Cref{fig:rewrite-judgments}, rule judgments & \leanfile{Ismr/Core/Rules.lean} & \leanin{Core}{Step}\\
\addlinespace
\multicolumn{3}{@{}l}{\textit{\Aref{app:interpretation-example}}}\\
\Cref{fig:interpretation-example}, the representations of $\Ctor B(0,1)$ & \leanfile{Examples/Worked.lean} & \leanin{Core}{Examples.Worked.b01_representations}\\
\addlinespace
\multicolumn{3}{@{}l}{\textit{\Aref{app:physical-codec}}}\\
\Aref{app:discriminant-closure}, discriminant closure & \leanfile{Ismr/Layout/Stored.lean} & \leanin{Core}{Stored, TagAccess, TagsStored}\\
Fuel-indexed codec & \leanfile{Ismr/Physical/Coding.lean} & \leanin{Core}{Physical.{encodeLayout, decodeLayout, encodeLayout_mono_le, decodeLayout_mono_le}}\\
\Cref{thm:encoding-existence}, encoding within capacity & \leanfile{Ismr/Physical/Direct.lean} & \leanin{Core}{Physical.encoding_exists_direct}\\
 & \leanfile{Ismr/Physical/Existence.lean} & \leanin{Core}{Physical.{ConcreteOn, AddressCapacity}}\\
\Cref{lem:storage-round-trip}, forward decoding & \leanfile{Ismr/Physical/Direct.lean} & \leanin{Core}{Physical.roundTrip_direct}\\
 & \leanfile{Ismr/Physical/Words.lean} & \leanin{Core}{Physical.Direct}\\
\addlinespace
\multicolumn{3}{@{}l}{\textit{\Aref{app:program-preservation}}}\\
Programs & \leanfile{Ismr/Programs/Syntax.lean} & \leanin{Core}{Programs.{Expr, Prog}}\\
\Cref{thm:derived-policies}, derived policies & \leanfile{Ismr/Core/Policies.lean} & \leanin{Core}{UpdateSeq.{decoder, decoder_decodes, observations}}\\
 & \leanfile{Ismr/Core/Inversion.lean} & \leanin{Core}{BodyUpdate, BodyUpdate.{undo, undo_decodes}, UpdateSeq, UpdateSeq.{undo, undo_decodes}, Ctxt.recon}\\
\Cref{lem:evaluation-correspondence}, evaluation correspondence & \leanfile{Ismr/Correctness/ProgramPreservation.lean} & \leanin{Core}{Programs.evaluation_value_iff}\\
\bottomrule
\end{longtable}
\endgroup

\endgroup

\end{document}